\documentclass[12pt,a4paper,epsf]{article}
\usepackage{graphics}
\usepackage{amssymb,amsmath}
\usepackage[dvips]{lscape,graphicx}
\usepackage{cite}
\usepackage[usenames]{color}

\definecolor{MyBlue}{rgb}{0.69,0.97,0.97}

\newcommand{\ct}{\cite}
\newcommand{\lb}{\label}

\newcommand{\bc}{\begin{center}}
\newcommand{\ec}{\end{center}}
\newcommand{\bd}{\begin{displaymath}}
\newcommand{\ed}{\end{displaymath}}
\newcommand{\be}{\begin{equation}}
\newcommand{\ee}{\end{equation}}
\newcommand{\ba}{\begin{array}}
\newcommand{\ea}{\end{array}}
\newcommand{\bea}{\begin{eqnarray}}
\newcommand{\eea}{\end{eqnarray}}
\newcommand{\bt}{\begin{tabular}}
\newcommand{\et}{\end{tabular}}
\newcommand{\un}{\underline}
\newcommand{\ov}{\overline}
\newcommand{\bp}{\begin{picture}}
\newcommand{\ep}{\end{picture}}
\newcommand{\bfi}{\begin{figure}}
\newcommand{\efi}{\end{figure}}

\begin{document}

\hyphenation{}

\begin{titlepage}



\centerline{\huge \bf {\textcolor{Blue}{
New Bound States of Top-anti-Top}}} 
\centerline{\huge \bf {\textcolor{Blue}{
Quarks and T-balls Production}}} 
\centerline{\huge \bf {\textcolor{Blue}{
at Colliders (Tevatron, LHC, etc.)}}\footnote{A talk given by
L.V.~Laperashvili at the seminar, ITEP, Moscow, April 2, 2008.}}

\vspace{2cm}

\centerline{\Large\bf C.D.~Froggatt${}^{1}$,
L.V.~Laperashvili${}^{2}$, R.B~Nevzorov${}^{1,2}$}

\vspace{5mm}

\centerline{\Large\bf H.B.~Nielsen${}^{3,4}$, C.R.~Das${}^{5}$}

\vspace{1cm}

\centerline{\itshape{\large ${}^{1}$ Department of
Physics and Astronomy,}}

\centerline{\itshape{\large Glasgow University,
Glasgow, Scotland}}

\vspace{5mm}

\centerline{\itshape{\large ${}^{2}$ A.I. Alikhanov Institute of Theoretical and
Experimental Physics,}}

\centerline{\itshape{\large Moscow, Russia}} 

\vspace{5mm}

\centerline{\itshape{\large ${}^{3}$ The Niels Bohr
Institute, Copenhagen, Denmark }}

\vspace{5mm}

\centerline{\itshape{\large ${}^{4}$ CERN CH 1211
Geneva 23, Switserland}}

\vspace{5mm}

\centerline{\itshape{\large ${}^{5}$ Center for High Energy
Physics, Peking University, Beijing, China}}


\thispagestyle{empty}
\end{titlepage}

\clearpage \newpage

\begin{abstract}
The present talk is based on the assumption that New Bound States
(NBSs) of top-anti-top quarks (named T-balls) exist in the
Standard Model (SM): a) there exists the scalar $1S$--bound state
of $6t+6\bar t$ --- the bound state of 6 top-quarks with their 6
anti-top-quarks; b) the forces which bind these top-quarks are
very strong and almost completely compensate the mass of the 12
top-anti-top-quarks forming this bound state; c) such strong
forces are produced by the interactions of top-quarks via the
virtual exchange of the scalar Higgs bosons having the large value
of the top-quark Yukawa coupling constant $g_t\simeq 1$. Theory
also predicts the existence of the NBS $6t + 5\bar t$, which is a
color triplet and a fermion similar to the t'-quark of the fourth
generation. We have also considered ''b-replaced'' NBSs: $n_b b +
(6t + 6\bar t - n_b t)$ and $n'_b b + (6t + 5\bar t - n'_b t)$,
etc. We have estimated the masses of the lightest ''b-replaced''
NBS: $M_{NBS}\simeq (300 - 400)$ GeV, and discussed the larger
masses of the NBSs. We have developed a theory of the scalar
T-ball's condensate, and predicted the existence of the three SM
phases, calculating the top-quark Yukawa coupling constant at the
border of two phases (with T-ball's condensate and without it)
equal to: $g_t \approx 1$. The searching for the Higgs boson H and
T-balls at the Tevatron and LHC is discussed.
\end{abstract}

\thispagestyle{empty}

\clearpage \newpage

\pagenumbering{arabic}

\bc \textcolor{Red} {\Large \bf Contents:}\ec

\vspace{0.3cm}

{\large \bf

\begin{itemize}

\item [1.]  Introduction: New colliders, the Higgs boson and
T-balls.

\item [2.] Higgs and gluon interactions of quarks.

\item [3.] T-balls' mass estimate.

\item [4.] The calculation of the top-quark Yukawa coupling
constant (YCC).

\item [5.] The main corrections to the calculation of YCC.

\item [6.] New phases in the SM.

\item [7.] Physical mass of the scalar $\Large \bf T_S$-ball.

\item [8.] Contributions of b-quarks in the "b-replaced NBS".

\item [9.] The Tevatron-LHC experiments searching for W, Z, t, t'
and different jets.

\item [10.] Can we see T-balls at LHC or Tevatron?

\item [11.] CDF II Detector experiment searching for heavy
top-like quarks at the Tevatron.

\item [12.] Charge multiplicity in decays of T-balls.

\item [13.] Conclusions.

\item [14.] Appendix. The Standard Model Lagrangian.

\end{itemize}}

\clearpage \newpage

\section{\un{Introduction: New colliders, the Higgs boson}\\ \un {and
T-balls.}}

The Salam-Weinberg theory of Electroweak (EW) interactions
describes very well the Standard Model (SM) which is confirmed by
all experiments of the world accelerators. This theory predicts
the existence of a scalar particle -- the Higgs boson. However,
this Higgs boson was not observed up to now in spite of the
careful searching for this particle. The main problem of the
future colliders: LHC, Tevatron, etc. -- is just the searching for
the Higgs boson H.

{\large \bf The Tevatron collider at Fermilab (Illinois, USA)
produces  the high energy collisions of proton-antiproton
 beams.}

Fermilab has been the site of several important discoveries that
have helped to confirm the SM of elementary particle physics.
Tevatron experiments observed the first evidence of the bottom
quark's existence (in 1977), and completed the quark sector of the
SM with the first observation of the top quark (in 1995).

{\large \bf Tevatron has the center-of-mass energy $\large \bf
\sqrt s = 1.96$ TeV, and therefore is currently the world's
highest energy particle collider.}

{\large \bf The Large Hadron Collider (LHC) is a new accelerator
being built at the European Organization for Nuclear Research
(CERN).}

The physics motivation provides the guidance for the construction
specifications of the LHC machine.

At the new frontier of the LHC of the High Energy Physics, the
areas that we aim to study with LHC can be summarized as follows.

\subsection{Explore the mechanism of the EW symmetry breaking.}

Although the SM of the EW interactions provides a successful
description of particles physics phenomenology (its predictions
have been verified by the experiments at LEP and Tevatron), the
mechanism of the EW symmetry breaking (EWSB) has not yet been
tested.

Within the SM, the EWSB is explained by the Higgs mechanism. {\bf
However, the mass of the Higgs boson is not predicted by the
theory.}

Direct searches in the previous experiments (mainly at LEP2) set a
low mass limit: $$\bf m_H \gtrsim 114.4 \,\,\,{\mbox{ GeV\,\,\, at
\,\,\, 95\% \,\,\, CL}}.$$ This limit can be indirectly
constrained from global fits to high precision EW data which
suggest a mass $$\bf m_H = 89^{+ 42}_{- 30}\,\,\,{\mbox{ GeV}}.$$
If we assume that there is not physics beyond the SM up to a large
scale $ \bf \Lambda$, then, on theoretical grounds, the upper
limit on the Higgs mass can be set to 1 TeV. Therefore, there is a
need for a machine that can probe the whole mass range, and LHC
has been designed for that.\\

The recent Tevatron result is:
$$ \bf 120 \lesssim M_H \lesssim 160\,\,\,{\mbox{GeV}}.$$

\subsection{Physics beyond the SM.}

There are several arguments which indicate that the SM is not the
final and complete theory. One of these, probably the strongest,
is the so-called {\large \bf hierarchy problem:} if the Higgs
particle exists, then the fermionic radiative corrections to its
mass will be described (at one-loop level) by the diagram of {\bf
Fig.~1}.

\bfi \centering
\includegraphics[height=50mm,keepaspectratio=true,angle=0]{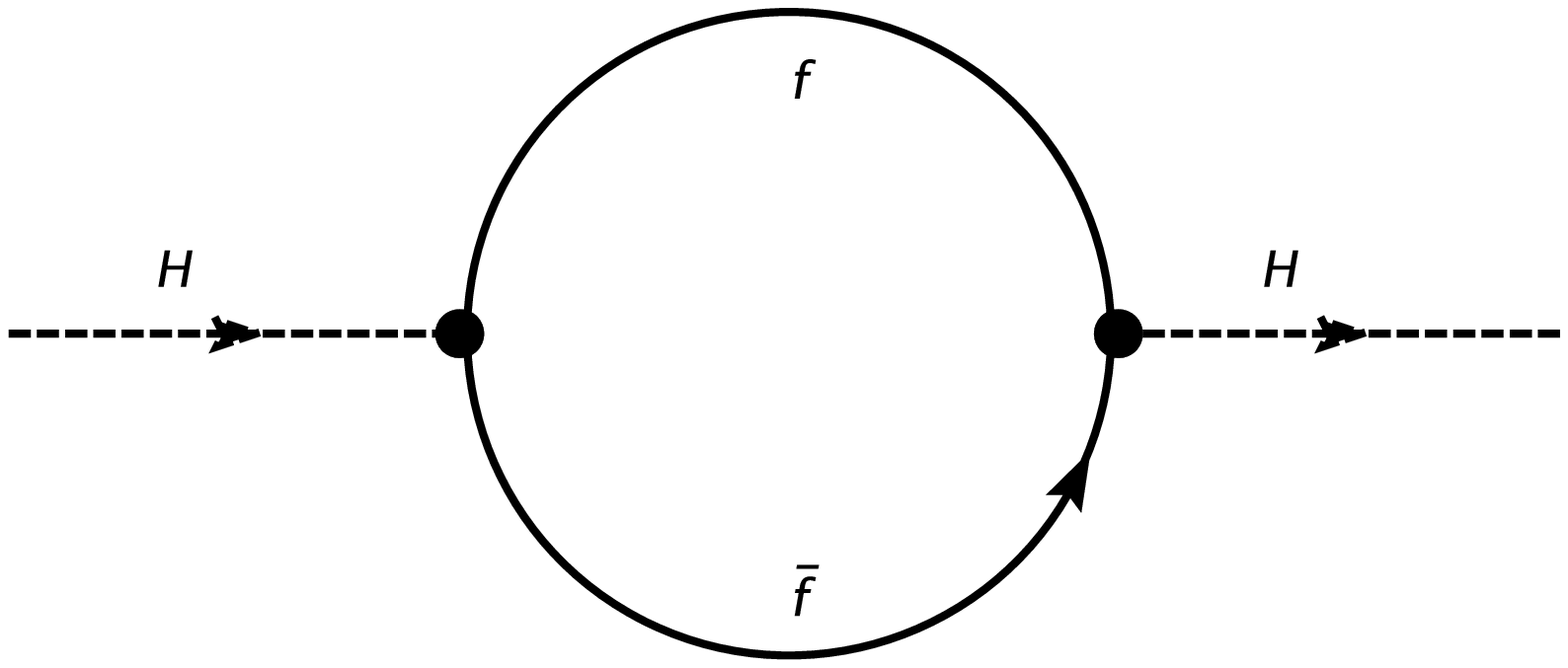}\caption{} \lb{1}\efi

Then the Higgs mass given by theoretical calculations depends on
the cut-off $\Lambda$ for the momentum in loop. Now, if there is
not new physics up to the Planck scale, then:
$$ \bf \Lambda \simeq M_{Planck}\simeq 10^{19}\,\,\, {\mbox{GeV,}}$$
and $$\bf M_H(renormalized) >> (1\,\,\,{\mbox{TeV}})^2,$$ unless
we fine tune so as to avoid that. Since the last value does not
agree with experimental limits, we start to believe that possible
new physics exists beyond the SM (for example, SUSY models, etc.).

\subsection{EW precision measurements.}

Because of the high energy and luminosity achieved, the  LHC will
be a factory of W and Z bosons, as well as of top and bottom
quarks.

\clearpage \newpage

It is estimated that the LHC, during the first year of operation,
will give the following events:
$$ \large \bf 10^8, \quad W \to e\nu,$$
$$ \large \bf 10^7, \quad Z \to e^+e^-,$$
$$ \large \bf 10^7, \quad t\bar t,$$
$$ \large \bf 10^{12}, \quad b\bar b.$$
LHC will establish the SM parameters. Any observed deviation from
the predicted values of the SM observables  will be a signal for
new physics.

The LHC is currently being constructed in the already existing LEP
tunnel of (approximately) 27 km circumference.

{\large \bf The machine will provide mainly proton-proton
collisions.} Also it will provide heavy ion collisions as well.

{\large \bf The LHC will produce two counter-rotating proton beams
with energy of 7 TeV each.

This gives 14 TeV center of mass energy ($\bf \sqrt s = 14$
TeV):\\

7 times bigger than the center of mass energy provided by Tevatron
at Fermilab!}\\

{\large \bf The completion of the LHC is expected at the end of
2008:

\bc at October, 21, 2008. \ec}

\subsection{New bound states of top-anti-top quarks.}

We hope that the LHC will provide a solution of the main puzzles
of EWSB. The present investigation is devoted to this problem and
based on the following three assumptions:\\

$\bullet$ there exists $\bf 1S$--bound state of $\bf 6t+6\bar t$,
e.g. bound state of 6 quarks of the third generation with their 6
anti-quarks;\\

$\bullet$ the forces which bind these top-quarks are so strong
that almost completely compensate the mass of the 12 top-quarks
forming this bound state.\\

$\bullet$ such strong forces are produced by the Higgs
interactions --- the interaction of top-quarks via the virtual
exchange by scalar Higgs bosons. They are determined by the large
value of the top-quark Yukawa coupling constant $\bf g_t$.\\

\clearpage \newpage

A new (earlier unknown) bound state $$\bf 6t+6\bar t,$$ which is a
color singlet, that is, ''white state'', was first suggested in
{\bf Ref.~[2] by Froggatt and Nielsen}
and now is named T-ball, or T-fireball.\\

$\bullet$ Theory also predicts the existence of {\bf the new bound
state $$\bf 6t + 5\bar t,$$ which is a fermion similar to the
quark of the fourth generation having quantum numbers of t-quark.}

The properties of T-balls are intimately related with the problem
of the Higgs boson observation.\\

The talk is based on the papers [1-9].

\clearpage \newpage

\section{\un{Higgs and gluon interactions of quarks.}}

If the Higgs particle exists, then between quarks $\bf qq$, quarks
and anti-quarks $\bf q\bar q$, and also between anti-quarks $\bf
\bar q\bar q$ there exist virtual exchanges by Higgs bosons (see {
\bf Fig.~2}). And in all these three cases we deal with attractive
forces.

 \bfi \centering
\includegraphics[height=50mm,keepaspectratio=true,angle=0]{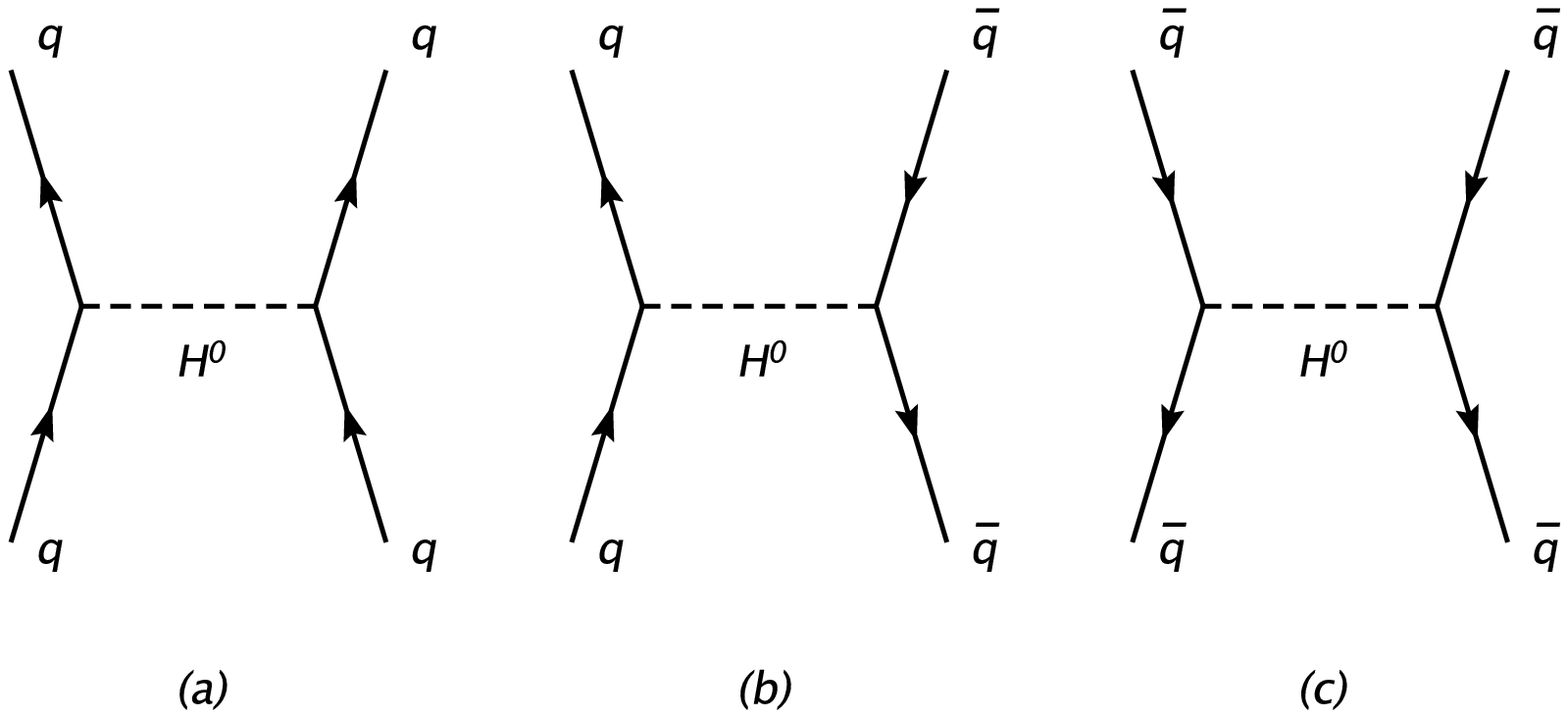}\caption{}\efi

It is well-known that the bound state $\bf t\bar t$ -- so called {
\bf toponium} -- is obligatory of the gluon virtual exchanges
(see { \bf Fig.~3}).\\

\bfi \centering
\includegraphics[height=50mm,keepaspectratio=true,angle=0]{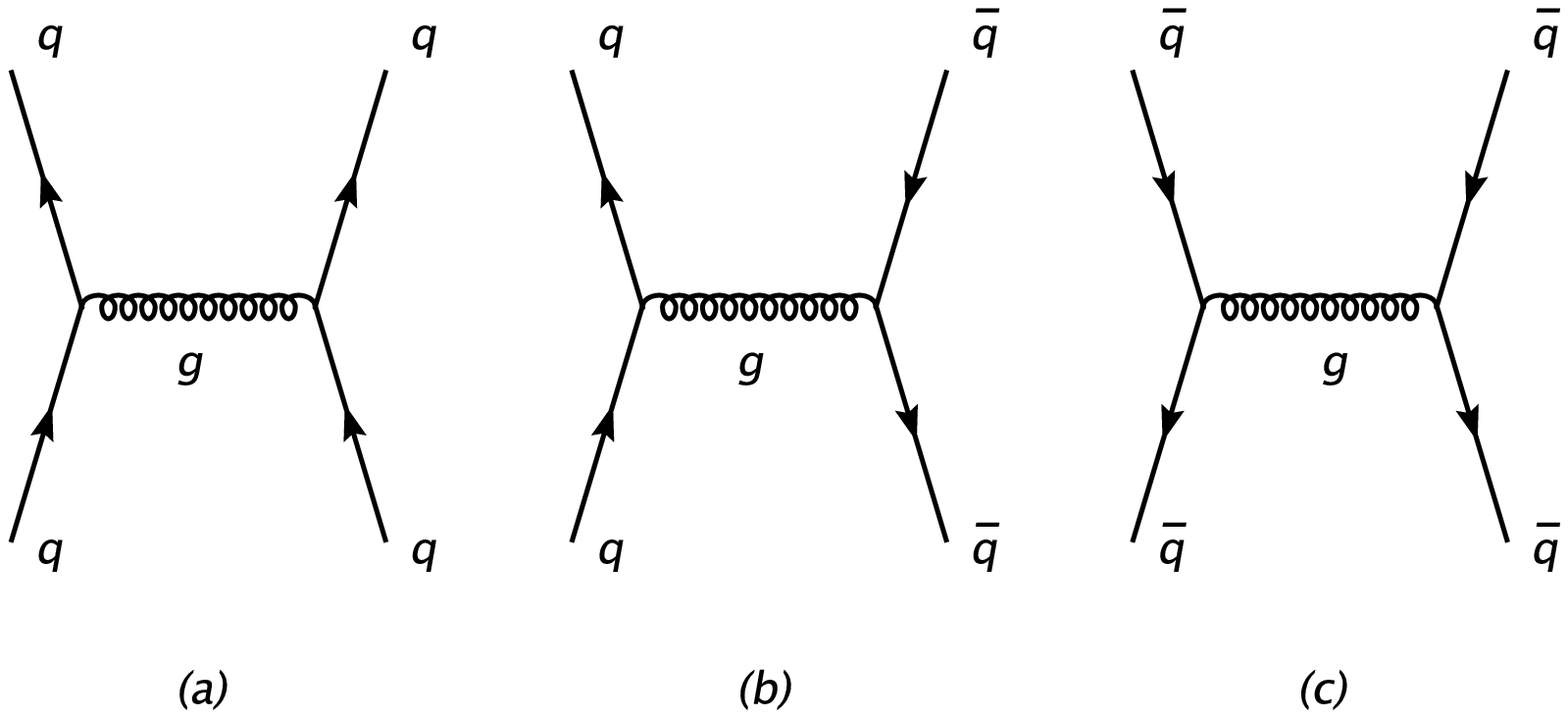}\caption{}\efi

In the case of toponium the contributions of the Higgs scalar
particles are essential, but less than gluon interactions.
Toponium is very unstable due to the decay of the top quark
itself. Had the latter been indeed stable it would have been a
very loosely bound state.

However, adding to the NBS more and more top and anti-top quarks,
we begin to notice that attractive Higgs forces increase and
increase. Simultaneously gluon (attractive and repulsive) forces
first begin to compensate themselves, and thus begin to decrease
relatively to the Higgs effect with the growth of the number of
NBS constituents $\bf t$ and $\bf \bar t$. The maximum of the
special binding energy value $\bf \epsilon$ (the binding energy
per top or anti-top) corresponds to the $\bf S$-wave NBS $\bf 6t +
6\bar t$. The explanation is given as follows: top-quark has two
spin states (two spin degrees of freedom corresponding to the two
projections of the spin $\bf \frac 12$) and three states of
colors. This means that, according to the Pauli principle, only $$
\bf 2\times 3=6 \,\,\, t-quarks$$ can create $\bf S1-wave\,\,\,
function$ together with $\bf 6\bar t-quarks$. So we deal with the
12 quark (or anti-quark) constituents, that is, with 6 pairs $\bf
t\bar t$), which simultaneously can exist in the $\bf S$-wave
state. If we try to add more top-constituents $\bf t\bar t$, then
some of them will turn out to the S2-wave, and the NBS binding
energy will decrease at least 4 times. For P-,D-, etc. waves the
NBS binding energy decreases more and more.

\section{\un{T-ball mass estimate [2-4].}}

T-ball mass containing the number $N_{const.}$ of top and anti-top
quarks is: \be  \bf
         M_T = N_{const.}M_t - E_T = N_{const.}(M_t -
 \epsilon)\,\,{\mbox{GeV}},
                                                \lb{1}  \ee
where $\bf M_t$ is the top-quark pole mass, $\bf E_T$ is a total
binding energy and $\bf \epsilon=E_T/N_{const.}$ presents the
specific binding energy.

Below we use the notation: {\bf the scalar NBS $\bf 6t + 6\bar t$,
having the spin $\bf S=0$, is named as $\bf T_s$-ball, and $\bf
T_f$-ball presents the NBS $\bf 6t + 5\bar t$, which is a fermion
($\bf \ov {T_f} = 5t + 6\bar t$).}

\subsection{$T_s$-ball mass estimate.}

According to the Particle Data Group \ct{10a}, the top-quark mass
is
$$ \bf M_t=172.6\pm 1.4\,\,\,GeV,$$ therefore the mass of the
$ \bf T_s$-ball is given by the following expression: \be \bf
         M_T = 12M_t - E_T = 12\cdot (172,6 - \epsilon)\,\, GeV.
                                                \lb{1a}  \ee
With aim to estimate the binding energy $ \bf E_T$ of the NBS $
\bf 6t+6\bar t$, first we will determine the binding energy of the
single top-quark relatively to the remaining 11 quarks, which we
shall call nucleus. Assuming, that the radius of this nucleus is
small enough in comparison with the Compton wave length of the
Salam-Weinberg Higgs particle, we are able to use the usual Bohr
formula for the binding energy of the Hydrogen atom, replacing the
electric charge $\large \bf e$ into the top-quark Yukawa coupling
constant (YCC) $ \bf g_t/{\sqrt 2}$.

Here we use the normalization, in which the kinetic energy term of
the Higgs field $ \bf \Phi_H$ and the top-quark Yukawa interaction
are given by the following Lagrangian density: \be  \bf
          L = \frac 12 D_{\mu}\Phi_H D^{\mu}\Phi_H + \frac{g_t}{\sqrt
          2}\ov{\psi_{tL}}\psi_{tR}\Phi_H  + h.c.   \lb{1b} \ee
In this case the attraction between the two top (anti-top) quarks
is presented by the potential similar to the Coulomb one: \be
 \bf         V(r) = - \frac{g_t^2/2}{4\pi r}. \lb{2a} \ee It
is easy to see that the attraction between any pairs $ \bf
tt,\,\,t\bar t,\,\,\bar t\bar t$ is described by the same
potential (\ref{2a}).

Now we can estimate the binding energy of a single top-quark
relatively to the nucleus having $ \bf Z=11$, using the well-known
equation  for the  $n$ energy level of the Hydrogen atom : \be
 \bf
    E_n = - \left(\frac{Zg_t^2/2}{4\pi}\right)^2\frac{M_t^{reduced}}{2n^2},
                                \lb{3a} \ee
where  $ \bf M_t^{reduced}$ is the top-quark reduced mass. Then:
\be  \bf
        M_t^{reduced} = \frac{ZM_t}{Z+1}, \lb{4a} \ee
and we obtain the following equation:
 \be  \bf
    E_n = - \left(\frac{Zg_t^2/2}{4\pi}\right)^2\frac{ZM_t}{2(Z+1)n^2}.
                                \lb{5a} \ee
The level with $ \bf n=1$ corresponds to the ground $ \bf S$-wave
state, e.g. \be  \bf  E_1 = -
\left(\frac{11g_t^2}{8\pi}\right)^2\frac{11M_t}{24}.
                                         \lb{5b} \ee
Here $ \bf g_t$ is the top-quark YCC.

In our normalization we obtain the following expression by the
Salam-Weinberg theory: \be  \bf M_t = \frac{g_t}{\sqrt 2}v \approx
174g_t \,\, GeV. \lb{5c} \ee

A total binding energy of $ \bf T_s$-ball, containing the 12
particles, can be obtained by adding the binding energy of the
remaining constituents, that is, by multiplying the formula
(\ref{5b}) with a general number of constituents, e.g. 12, taking
into account a duplication.

Finally, in this {\large \bf \un{non-relativistic case}} the value
of the total binding energy is equal to: \be  \bf
    E_T = 6\left(\frac{11g_t^2}{8\pi}\right)^2\frac{11M_t}{24} =
             \left(\frac{11g_t^2}{4\pi}\right)^2\frac{11M_t}{16}.
                                \lb{6a} \ee
However, by analogy with a hydrogen-like atom, we have considered
only $\large \bf t$-channel exchange by the Higgs bosons between
the two top (or anti-top) quarks in the system of the NBS.

Let us consider now $\bf u$-channel exchange.

From the first point of view, it is expected the absence of the
difference between the quarks of different colors. But if we
consider a formalism, in which both degrees of freedoms (colors
and spin states of quarks) are fixed, then the NBS $\bf 6t+6\bar
t$ is completely antisymmetric under the permutation of its color
and spin states. In this case, we can easily estimate $\bf
u$-channel contributions. Assuming that the NBS is antisymmetrized
in such a manner, we formally consider a quark as a particle
having no degrees of freedom. In this case, we shall take into
account ''minus'' under the permutation of two quarks. It is
natural, that in this approach a quark plays a role of a boson,
but not a fermion.

As a result of s-, t-, and u-channel exchanges, we have the
following expression for the total binding energy:
 \be  \bf
    E_T = \frac{33g_t^4}{(4\pi)^2}\cdot 12M_t.
                                \lb{7b} \ee
Considering a set of Feynman diagramms (e.g. the Bethe-Salpeter
equation), we obtain the following Taylor expansion in $\bf g_t^2$
for the mass of the NBS $\bf T_s$, containing 12 top-anti-top
quarks:
$$  \bf
    M_T^2 = (12M_t)^2 -2(12M_t)E_T + ...$$
     \be  \bf  =
    (12M_t)^2(1 - \frac{33}{8\pi ^2}g_t^4 + ...).
                                \lb{8a} \ee

\subsection{$T_f$-ball mass estimate.}

One of the main ideas of the present investigation is to show that
the Higgs interaction of the 11 top-anti-top quarks creates a $
\bf T_f$-ball -- {\large \bf the new fermionic bound state $\large
\bf 6t + 5\bar t$, which is similar to the quark of the fourth
generation with quantum numbers of the top quark.} We have tried
to estimate the $\bf T_f$-ball's mass.

In general, the binding energy of the top-quark in the NBS depends
on the number of the NBS constituents $\bf N_{cost.}$, and is
proportional to the following expression: \be  \bf
      E_{binding}\propto \frac 12 N_{const.}(N_{const.} - 1).
      \lb{13} \ee
The dependence of the T-ball's mass of $\bf N_{cost.}$ is given by
{\bf Fig.~4}.

\clearpage \newpage

\bfi \centering
\includegraphics[height=100mm,keepaspectratio=true,angle=0]{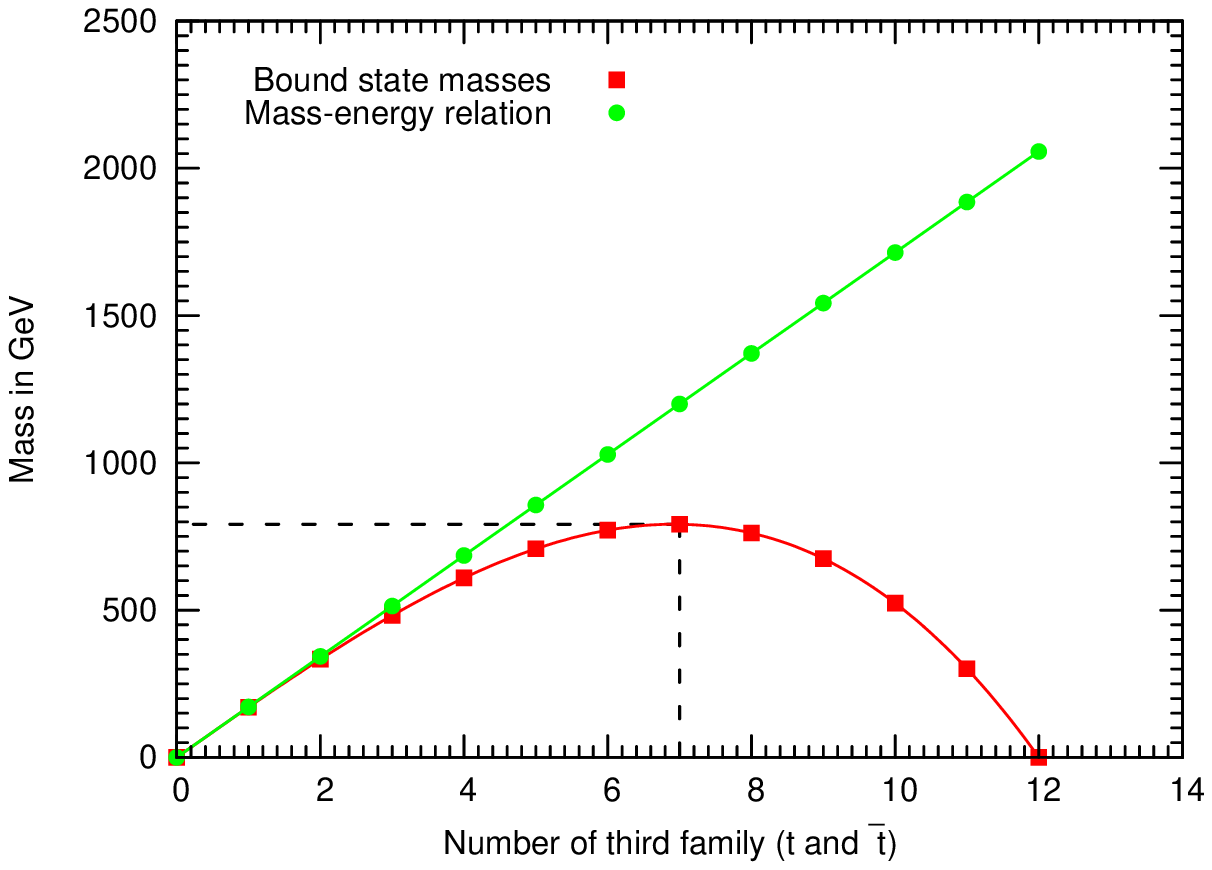}
\caption{The dependence of the T-ball's mass of the number $\bf
N_{cost.}$ of the NBS constituents. } \lb{4}\efi

\clearpage \newpage

This dependence is described by the following equation: \be \bf
   M_T =  M_{NBS} = M_t\cdot N_{const.}(1 - \frac{N_{const.}^2}{12^2}).
       \lb{14} \ee
According to the formula (\ref{14}), we have obtained the estimate
of the $\bf T_f$-mass of the NBS $\bf 6t + 5\bar t$, using
Particle Data Group result $ \bf M_t = 172,6\pm 1.4$ GeV:
$$\bf M_{T_f}\approx (172,6)\cdot 11\cdot 0.16 \,\,\,GeV
\simeq 300\,\,\, GeV.$$ It is necessary to notice the increasing
of the mass of $\bf T_s(b-replaced)$-ball, which is formed by the
replacement of a t-quark by a b-quark in $\bf T_s$-ball (see
Ref.~\ct{7} and Section 8):
$$\bf T_s(b-replaced) = 5t + b + 6\bar t,$$
in comparison with mass of $\bf T_s$.

It is obvious that $$\bf M_{T_s(b-replaced)} > M_{T_s},$$ giving
$$\bf M_{T_s(b-replaced)}\backsimeq M_{T_f}.$$ It is
obvious that considering the different T(b-replaced)-balls we can
obtain more heavy T-balls:
$$\bf M_{T(b-replaced)} \gtrsim 400\,\,\,GeV.$$

\section{\un{The calculation of the top-quark YCC}\\\un {at the two phases border.}}

According to the Salam-Weinberg theory (SM), we have Eq.~
(\ref{5c}), from which using the experimental value of the
top-quark mass \ct{10a} $$ \bf  M_t\approx 172.6\pm 1.4
\,\,\,{\mbox{GeV}},$$ we obtain: \be  \bf
      g_t\approx \frac{M_t}{v/{\sqrt 2}}
      \approx (172.6\pm 1.4)/174.5 \approx 0.989 \pm 0.008,     \lb{4} \ee
that is, top-quark YCC is of order of unity at the EW-scale.

At present, a lot of investigators, theorists and
experimentalists, are looking forward to the New Physics. And it
is quite possible that the {\large \bf ''Bjorken-Rosner
nightmare''} will take place: LHS will discover the Salam-Weinberg
Higgs boson and nothing more. Nevertheless, the NBS T-ball can
exist, because it is calculated in the framework of the SM.
Supersymmetry, for instance, cannot exclude this phenomenon: only
can change the details of calculations.

The light scalar Higgs bosons can bind top-quarks so strongly that
finally we shall obtain {\large \bf the Bose-condensate of T-balls
in the vacuum}, in which we live, e.g. in the EW-vacuum. Indeed,
it is quite possible: for example, if $\bf g_t$ increases when the
number of top-quarks in T-ball grows, then the binding energy
compensates the NBS mass $\bf 12 M_t$ in the $\bf T_s$-ball
(having $\bf 6t+6\bar t$) so strongly that the mass $\bf M_{T_s}$
becomes almost zero, and even tachyonic, e.g. $\bf M_{T_s}^2 < 0$,
what means the formation of the scalar T-balls' condensate in the
vacuum. The result $\bf g_t\sim 1$ means that the experimentally
observed value of the top-quark YCC belongs to the
border of two phases -- phase-I and phase-II:\\

I) the phase-I has no the Bose-condensate of T-balls,\\

II) but the phase-II contains such a condensate.\\

In this case the effective potential $\bf V_{eff}(|\Phi_T|)$,
depending on the norm of the T-ball scalar field $\bf \Phi_T$, is
presented by {\bf Fig.~5}.

\bfi \centering
\includegraphics[height=100mm,keepaspectratio=true,angle=0]{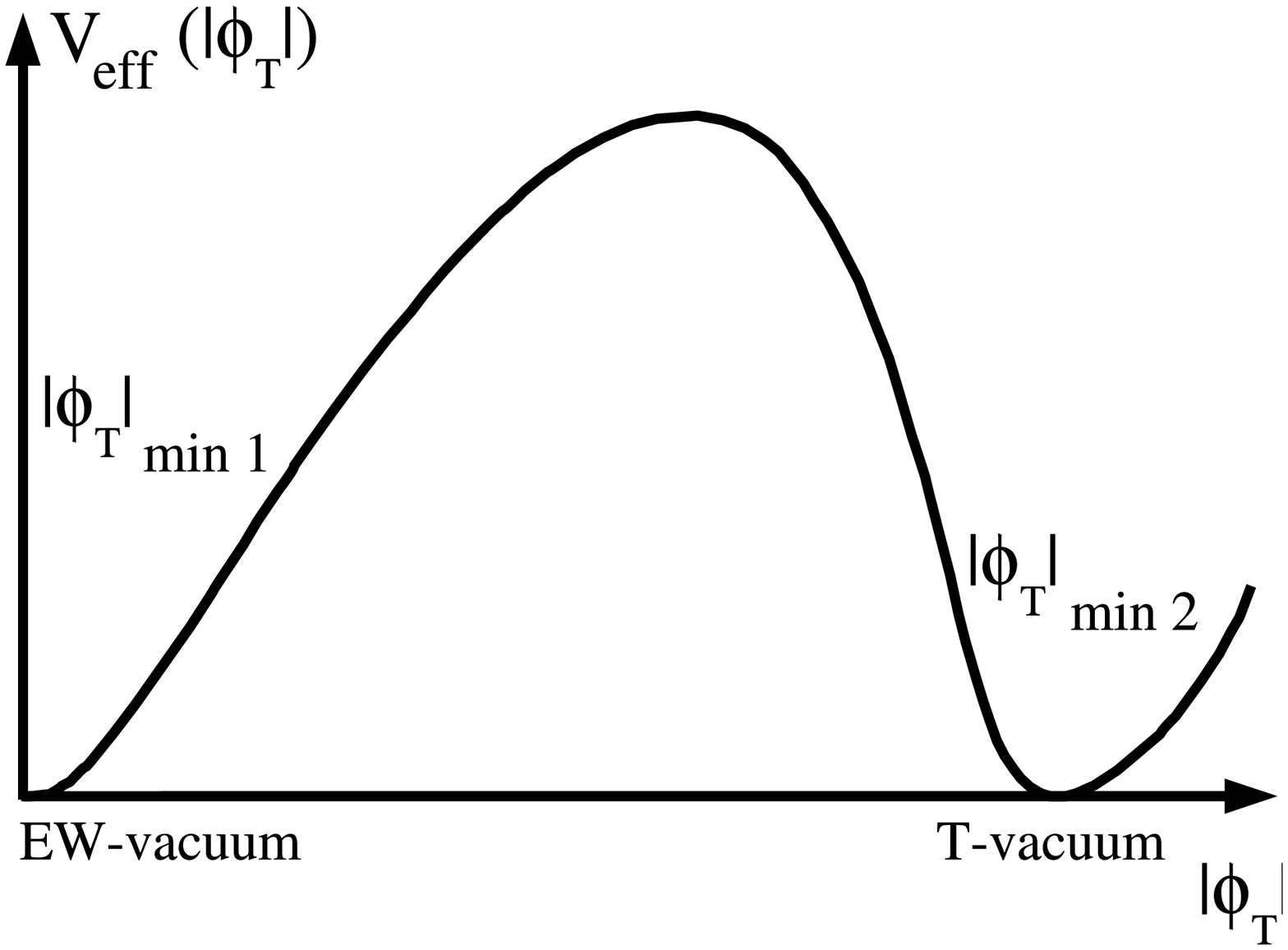}
\caption{The effective potential $\bf V_{eff}(|\phi_T|)$,
depending on the norm of the T-ball scalar field $\bf \phi_T$, has
two minima: at $\bf \phi_T = 0$ (EW-vacuum) and at $\bf \phi_T
\neq 0$ (T-vacuum). }\efi

\clearpage \newpage

We see that the main requirement of the appearance of the new
phase of the condensed $\bf T_s$-balls is a condition:
$$\bf m^2_{NBS} = M_{T_s}^2 = 0.$$
Using Eq.~(\ref{8a}), which describes the square mass of the
scalar fireballs, it is easy to obtain the estimate of the YCC
value of top-quark at the border of the two phases I and II
\ct{3}: \be \bf {g_t|}_{p.t.b.}\equiv
g_t|_{phase\,\,transition\,\, border} \approx (\frac{8\pi
^2}{33})^{1/4}\approx 1.24. \lb{9a} \ee However, there is an
additional problem.

The fluctuations of the Higgs field $\bf \Phi_H$ insight the NBS $
\bf T_s$ become stronger and stronger when YCC $\bf g_t$ increases
and the NBS radius decreases. As a result, the mean value of the
Higgs field can become negative, in comparison with its vacuum
value. Taking into account the configuration of the top-quark
Dirac sea insight the NBS, we see that in this case the Higgs
field with an opposite sign can become a vacuum value. Such a
Higgs field configuration is described by the situation when the
non-relativistic kinetic term for quarks together with the mass
energy of the NBS are equal to zero (at least approximately). The
estimate of such fluctuations were obtained in the paper \ct{3}
and gave the following result: \be  \bf {g_t|}_{p.t.b.} = 1.06\pm
0.18, \lb{10a} \ee which is in agreement with the experimental
value (\ref{4}) obtained at the EW-scale. With b-quarks
contributions {see Ref.~\ct{7} and Section 8) we have:
 \be  \bf g_t|_{(p.t.b.)}\approx \sqrt{(1/2)}\cdot 1.24 \approx 0.87_7.
       \lb{11a}  \ee
The calculation of accuracy, given below and equal to 8.5\%, gives
the following result: \be  \bf
   g_t|_{(p.t.b.)}\approx 0.88\pm 0.07. \lb{12a}  \ee

\section{\un{Main corrections to the top-quark YCC calculation.}}

What is the main corrections to the value of the top-quark YCC
given by Eq.~(\ref{11a}) at the border
of the two phases I and II ?\\

As will be discussed below in Section 8, we first take into
account that there virtually will be $\bf b\bar{b}$ pairs
replacing the top pairs, but only the left handed components can
come in (see Ref.~\ct{7}). On top of that we then have the
following minor corrections listed:

1) The first correction comes from gluon interactions if we take
into account simultaneously the Higgs and gluon interactions of
top-quarks in all (s-, t-, u-) channels.

2) The correction from the one-loop interaction of top-quarks.

3) The correction due to that the effective Higgs mass $\bf m_H$
is not zero - as we first calculate with - but rather varies as a
function of the distance $\bf r$ from the center, first reaching
the normal effective Higgs mass value --- say, the LEP finding
value $\bf m_H \cong 115$ GeV --- in the outskirts of the T-ball.

4) Relativistic corrections.

5) Renormgroup corrections.

6) The corrections from manybody effects --- from the
contributions of not only one-, but n-Higgs-bosons.

In general, all these corrections lead to {\bf the accuracy 8.5\%}
and give the result (\ref{12a}).

As it was shown in papers \ct{2,3,4}, the further increasing of
YCC $\bf g_t$ can give:
            $$ \bf M_{T_s}^2 < 0, $$
and T-balls begin to condense, forming a new phase of the SM ---
the phase of the condensed T-balls.

\section{\un{New phases of the SM.}}

Now we are in confrontation with a question: do the new phases of
the SM exist? Are they different from the well-known
Salam-Weinberg Higgs phase? Does a phase of the condensed T-balls
exist?

The answer on this question is related with the SM parameters.

\subsection{Three EW phases of the SM.}

Taking into account seriously our results in the estimates of $\bf
g_t$ and $\bf M_T$, we can have three phases -- three vacua of
the SM at the EW-scale:\\

I) $<\Phi_H> \neq 0 , <\Phi_T> = 0$ --- ''Vacuum 1'', the phase
in which we live;\\

II) $<\Phi_H> \neq 0 , <\Phi_T> \neq 0$ --- ''Vacuum 2'' (honestly
speaking it is a bit speculative, because it is also possible that
in the $<\Phi_T>$-condensate phase $<\Phi_H> =0$):\\

III) $<\Phi_H> = 0$,
$<\Phi_T> \neq 0$ --- ''Vacuum 3'',\\

which are presented symbolically by the phase diagram of {\bf
Fig.~6}.

\clearpage \newpage

\bfi \centering
\includegraphics[height=120mm,keepaspectratio=true,angle=0]{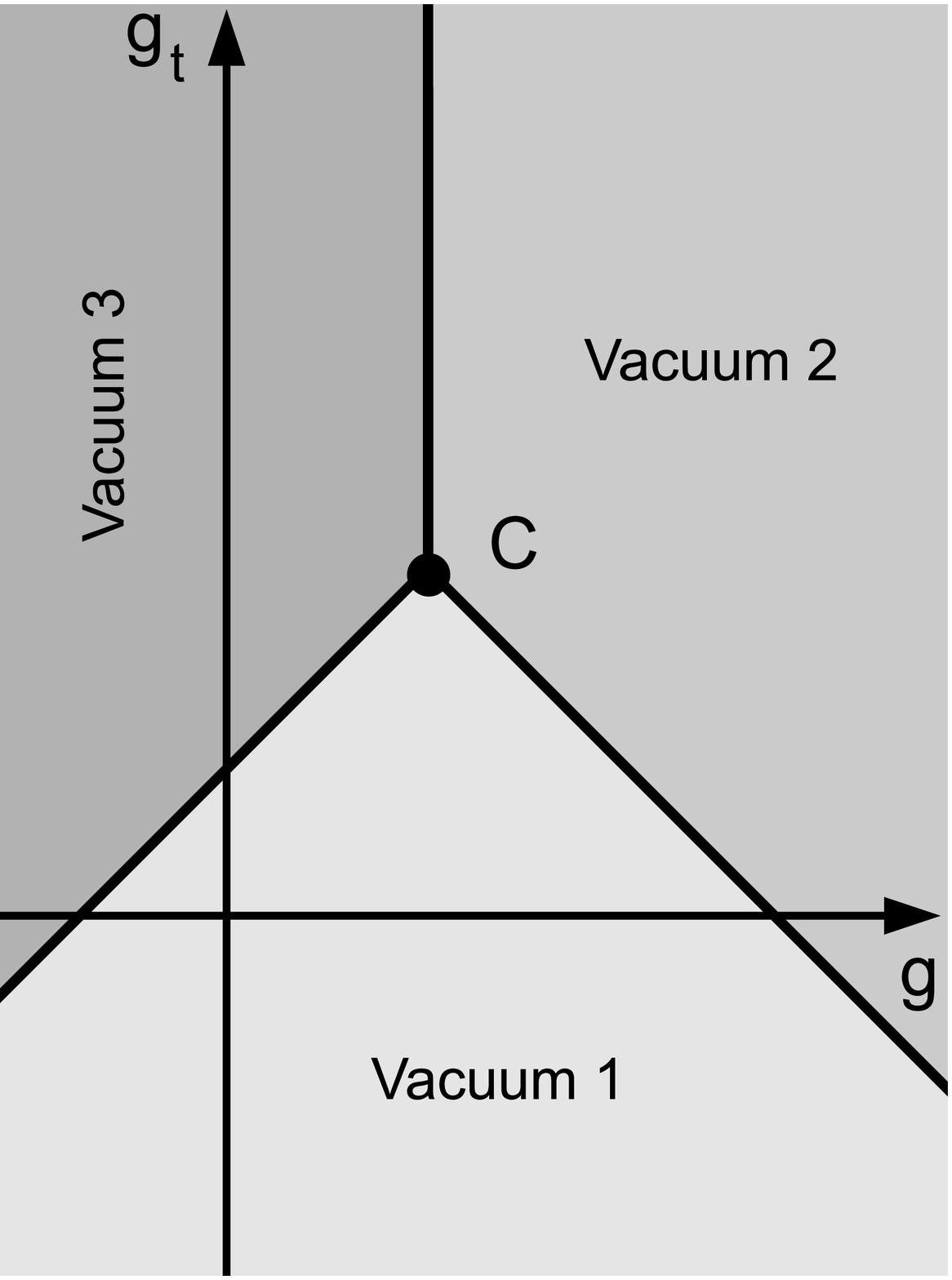}\caption{A symbolic phase diagram for the SM at the EW-scale.}\efi

\clearpage \newpage

{\bf Fig.~6} shows the critical point C (triple point), in which
three phases meet together: this triple point is similar to the
critical point considered in thermodynamics where the density of
the vapor, water and ice are equal (see {\bf Figs.~7,8}).

\bfi \centering
\includegraphics[height=60mm,keepaspectratio=true,angle=0]{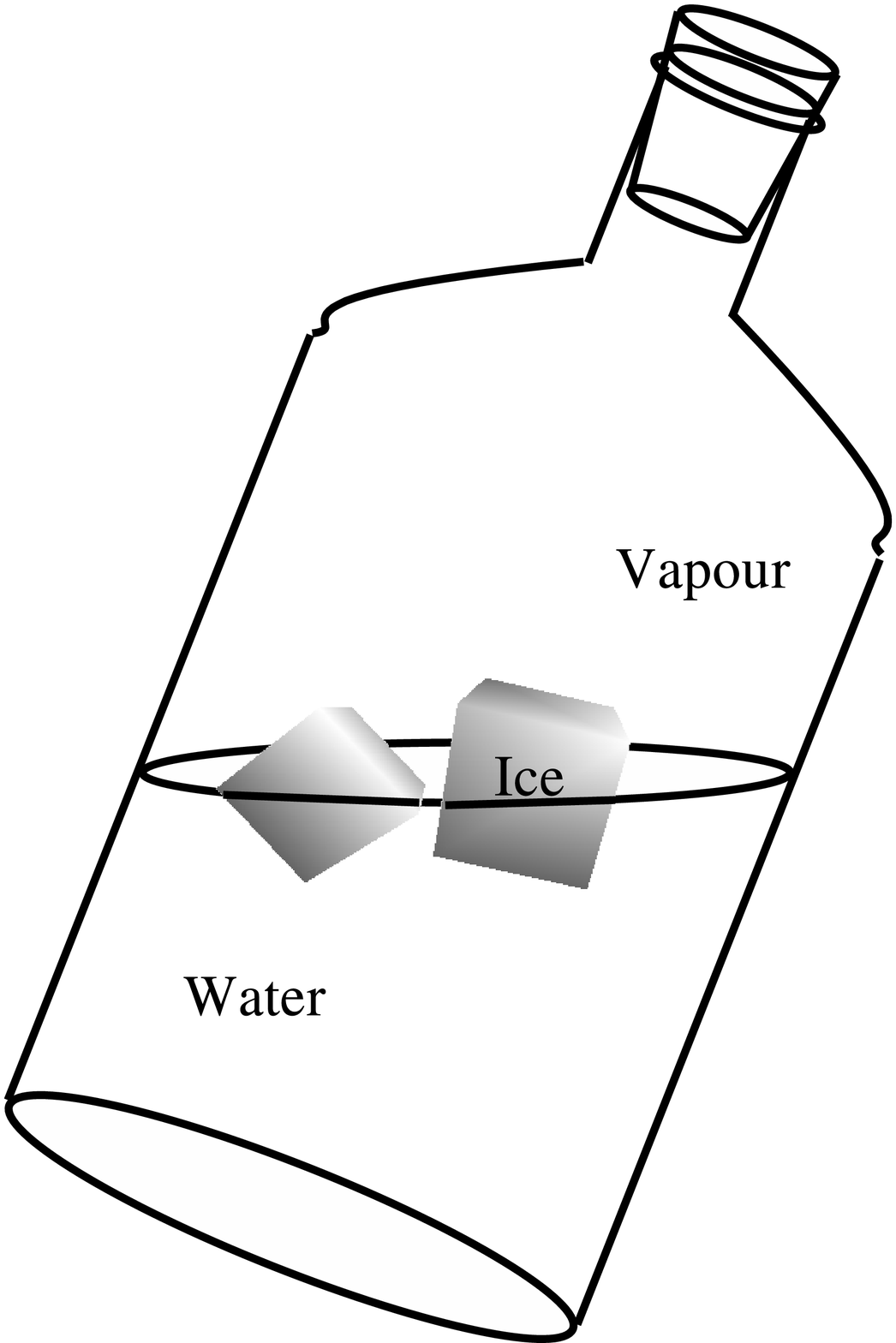}\caption{}\efi

\bfi \centering
\includegraphics[height=60mm,keepaspectratio=true,angle=0]{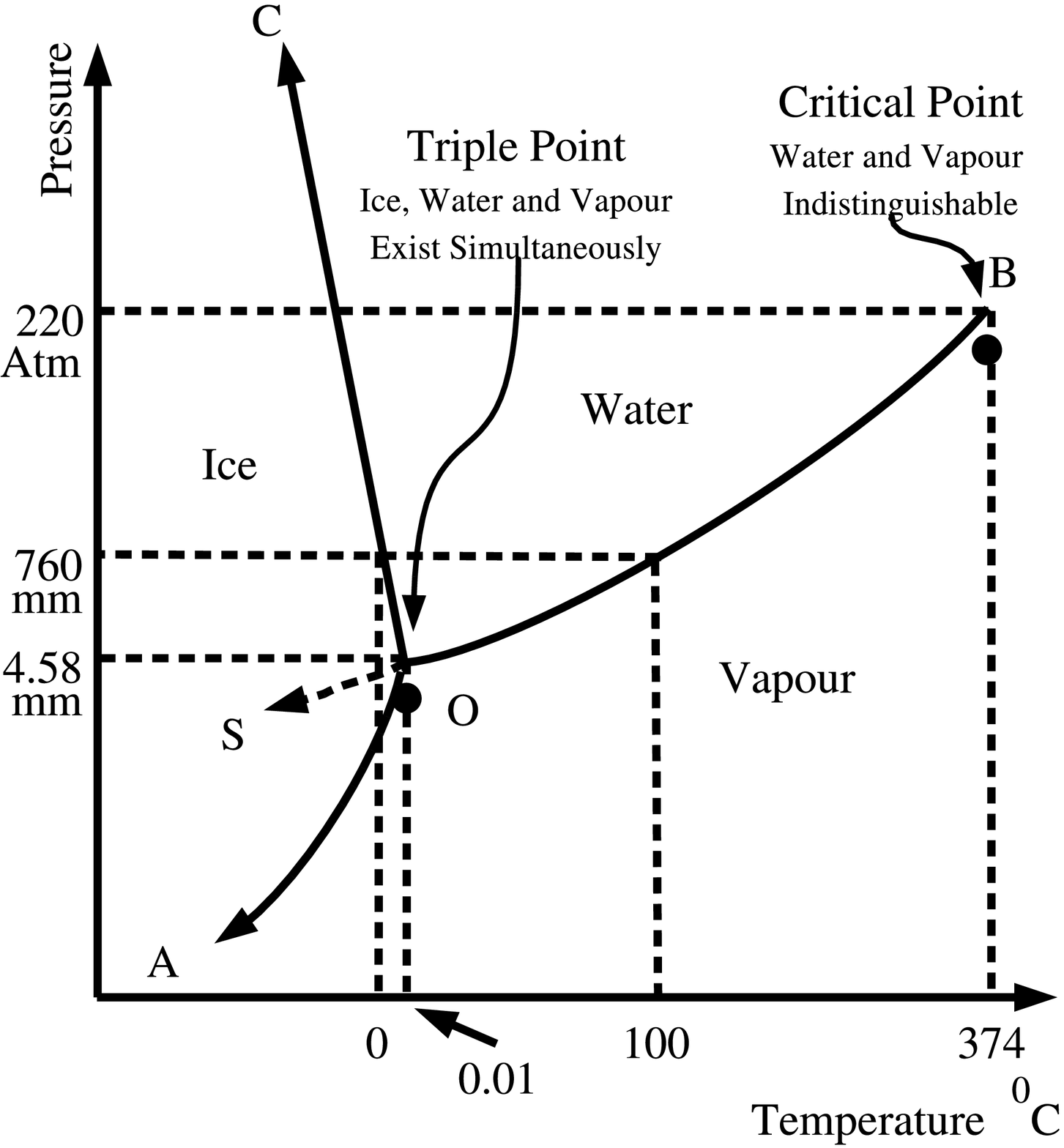}\caption{}\efi

The existence of the new phases near the EW-scale can solve the
problem of hierarchy. Here we recall {\un{ \bf The Multiple Point
Principle (MPP)}}, suggested in \\
Refs.~\ct{10,11,12,13,14,15,16,17}.

The calculation of the NBS mass have used only the SM parameters.
The MPP determines the coupling constants in the SM and therefore
--- the structure of the NBSs $\bf T_{s,f}$. Since at the border of
the two phases I and II the top-quark YCC leads to zero mass of
the NBS $\bf T_s$, we can assume that the MPP manifests the phase
transitions in the SM in such a way that we have the finetuning in
the SM, which solves the hierarchy problem.

\subsection{The fundamental (Planck) scale of the SM. }

{\it A priori} it is quite possible for a quantum field theory to
have several minima of its effective potential as a function of
its scalar fields $\bf \Phi$ (exactly speaking of its norm $ \bf
|\Phi|$). These minima can be degenerate. Moreover, it is assumed
that all vacua existing in Nature (there can be a number of
several vacua) are degenerate and have the same zero, or almost
zero, vacuum energy densities which coincide with the cosmological
constant $\Lambda$ determined by Einstein. This is confirmed by
the phenomenological Cosmology.

According to the MPP, the SM has the two minima of its effective
potential considered as a function of the variable $ \bf
|\Phi_H|$. These minima are degenerate and have $\bf \Lambda = 0$:
\be
 \bf              V_{eff}|_{min1}= V_{eff}|_{min2}=0,
\lb{11} \ee \be  \bf
             V'_{eff}|_{min1}= V'_{eff}|_{min2}=0,   \lb{12}
\ee what is shown in {\bf Fig.~9}.

\bfi \centering
\includegraphics[height=80mm,keepaspectratio=true,angle=0]{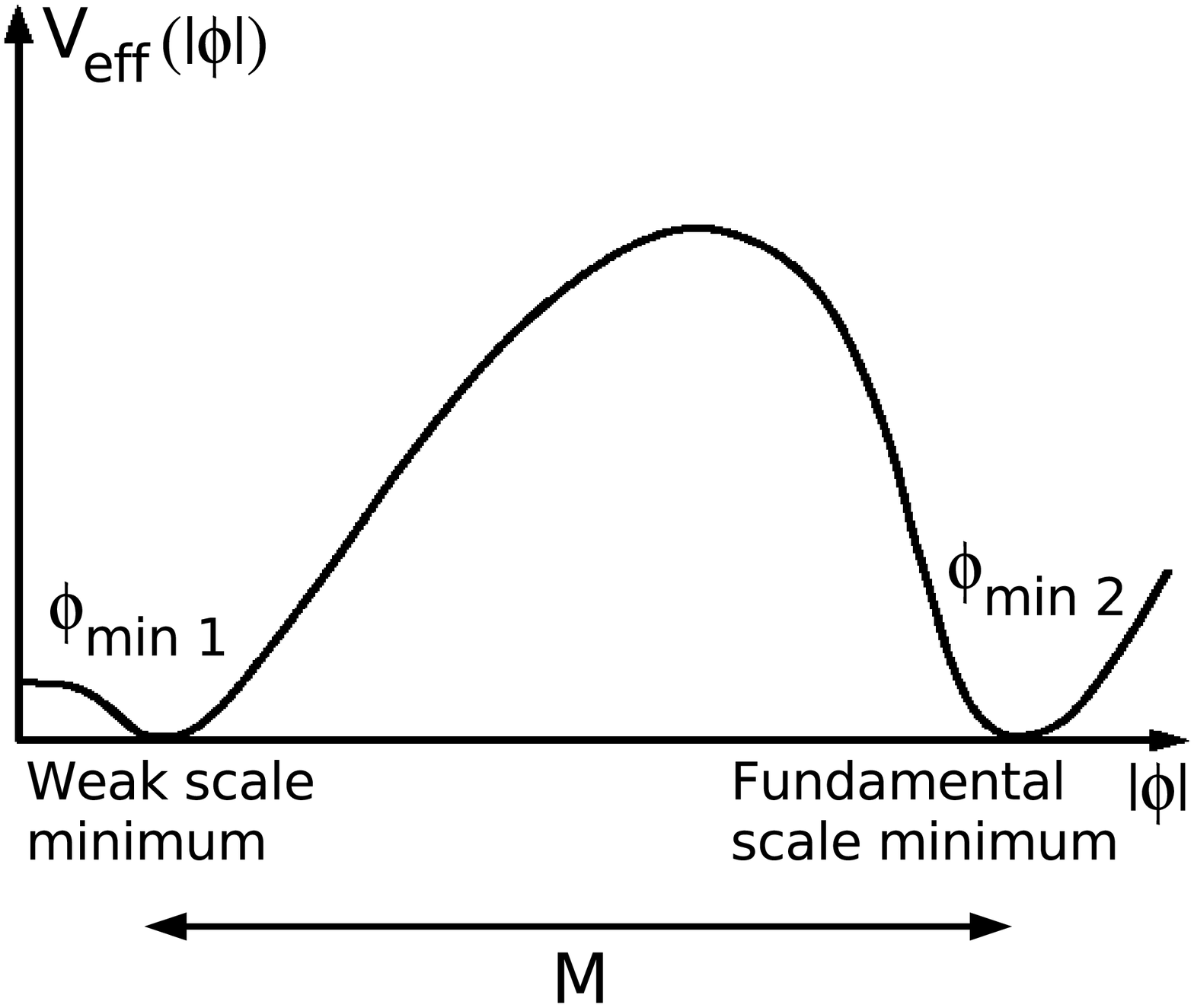}
\caption{The first (our) vacuum at $|\phi|\approx 246$ GeV and the
second vacuum at the fundamental scale $|\phi|\sim M_{Pl}$.}\efi

It is assumed that the second minimum exists near the Planck
scale:
$$\bf |\Phi_{min2}|\sim M_{Pl}.$$
This scale is considered as a fundamental one.

\clearpage \newpage

\section{\un{Physical mass of scalar $T_S$-ball.}}

If we have a condensate of scalar $\bf T_s$-balls with the mass
$\bf m_{NBS}$, then we can consider the potential similar to the
Higgs one. In general, we have the potential: \be  \bf U = \frac
12 m_{NBS}^2|\Phi_{NBS}  |^2 + \frac{\lambda_s}{4}|\Phi_{NBS}|^4 +
C, \lb{1y} \ee where C is a cosmological constant.

If NBS is a tachyon, then $ \bf m_{NBS}^2= - \mu^2$, and we have a
condensate of $ \bf T_s$-balls given by the second vacuum of { \bf
Fig.~5}, when: $$ \bf  U' = 0.$$ This condensate is described by
the field: \be  \bf <|\Phi_{NBS}|> = \frac{\mu}{\sqrt{\lambda_s}}.
\lb{2y} \ee Now we are able to determine a physically existing
scalar NBS which can be observed experimentally. This NBS is
similar to the fundamental particle, which can be described by the
field $ \bf \Phi_{phys.NBS}$.

Its mass $\bf m_{phys. NBS}$ is determined by the following
requirement:
 \be  \bf \frac{\partial^2 U}{\partial |\Phi_{NBS}|^2} =
m^2_{phys. NBS}.
      \lb{3y} \ee
Calculating: \be  \bf  \frac{\partial U}{\partial |\Phi_{NBS}|} =
m_{NBS}^2 |\Phi_{NBS}| + \lambda_s |\Phi_{NBS}|^3, \lb{4y} \ee we
obtain: \be  \bf \frac{\partial^2 U}{\partial |\Phi_{NBS}|^2} =
m_{NBS}^2 + 3\lambda_s |\Phi_{NBS}|^2.  \lb{5y} \ee Taking into
account Eqs.~(\ref{2y}) and (\ref{5y}), we obtain: \be  \bf
\frac{\partial^2 U}{\partial |\Phi_{NBS}|^2} = -\mu^2  +
3\lambda_s\cdot \frac{\mu^2}{\lambda_s} = 2\mu^2 = m^2_{phys.
NBS}. \lb{6y} \ee Then the mass of this physical scalar NBS $ \bf
T_S\equiv T_s(phys.)$ is \be  \bf  M^2_{T_S} = m_{phys.NBS}^2 =
2\mu^2. \lb{7y} \ee This particle is not tachyon already. This is
a scalar particle. Its mass will be estimated later.

\section{\un{Contributions of b-quarks in the "b-replaced NBS".}}

Up to the year 2008, we were sure that only $\bf t$- and $\bf \bar
t$-quarks are the constituents of T-balls. But at present we know
that we can take into account considerable contributions of left
b-quarks (see Ref.~\ct{7}).

If we had no $\bf b-\bar b$-quark pairs in T-balls, then there
would be an essential superposition of different states of the
weak isospin. The presence in the condensate of the not pure
singlet states of the EW-theory could create serious problems, so
it is better to live in the vacuum phase-I without T-ball
condensate having
$$ \bf <\Phi_T>=0.$$
But the presence of b-quarks in the NBS leads to the dominance of
the isospin singlets of EW-interactions only, and even that we
should live in the phase-II, it can be considered without any
problems, w.r.t. agreement with the SM LEP precission data.

With the inclusion of both b and t quarks we think of a more weak
isospin invariant picture, and it becomes natural to think of
replacing one (or several) of the t-quarks in the NBSs by
b-quark(s).

Now the ``b-replaced'' scalar NBS still would have a mass very
close to ''11'' NBS $\bf M_{T_f}$, say, of the order of 400 GeV,
by our estimate. It is a boson: \be  \bf T_S(b-replaced) = b + 5t
+ 6\bar t. \lb{B1} \ee We have also: \be \bf T_S(\bar b-replaced)
= 6t + \bar b + 5\bar t. \lb{B2} \ee Of course, we also can
consider the fermionic b-replaced NBS: \be
 \bf T_f(b-replaced)= b + 5t + 5\bar t, \lb{B3} \ee and \be
 \bf \ov{T_f}(\bar b-replaced)= 5t + 5\bar t + \bar b.
\lb{B3a} \ee In general case we can construct: \be  \bf
T_S(n_bb-replaced) = n_b b + (6t + 6\bar t - n_b t), \lb{B4} \ee
and \be \bf T_S(n_{\bar b}\bar b-replaced) = n_{\bar b}\bar b +
(6t + 6\bar t - n_{\bar b}\bar t). \lb{B5} \ee Correspondingly we
can obtain: \be \bf T_f(n_bb-replaced) = n_b b + (6t + 5\bar t -
n_b t), \lb{B6} \ee and \be  \bf \ov{T_f}(n_{\bar b}\bar
b-replaced) = n_{\bar b}\bar b + (5t + 6\bar t - n_{\bar b}\bar
t). \lb{B7} \ee

\subsection{The important estimate of the mass of the ``b-replaced NBS''.}

There is a simple way to estimate the mass of the NBS with one t
replaced by a b, what we called ``b-replaced NBS'': $ \bf
T_f(b-replaced) = 5t + b + 5\bar t$, or $ \bf T_f(b-replaced,
b\bar b) = 5t + b + n_b b\bar b + 5\bar t$, etc.

We have seen that the b does not interact significantly (in the
first approximation) with the NBS. Thus we can add a b-quark (or
anti-b-quark) to the NBS ''11'' without changing the energy or
mass. Then the b-replaced scalar NBS will have a mass $\bf
\backsimeq 300$ GeV. And the balls $\bf T_f(b-replaced) = 5t + b +
5\bar t$ and $\bf T_f(b-replaced, b\bar b) = 5t + b + n_b b\bar b
+ 5\bar t$ will have a mass $\bf > 300\,\,{\mbox{GeV}}$. We can
consider more heavy T-balls with $\bf M_T > 400 \,\,{\mbox{GeV}}$,
but they will have smaller cross-sections of their production,
because they are less strongly bound and can a less extend to be
considered approximately fundamental particle.

\subsection{Two-gluon diagrams for the NBS production.}

The ``b-replaced NBS'' $\bf T_S(b-replaced)$ cannot be produced
simply in a pair by a gluon vertex, because it is a color singlet
$\bf \un{1}$.

A pair $\bf T_f\ov{T_f}$ of ``11'' NBS  can be produced by a
gluon, because it is a color triplet, and then ``11'' could make a
rather soft emission of a b-quark and also a scalar ``b-replaced
NBS''. But such a ``soft b'' emission may be difficult to detect
at FNAL (Tevatron).

There also is an alternative idea (see Ref.~\ct{8}).

According to the idea by Li-Nielsen, the t or b emission and the
scalar ``b-replaced NBS'' might be produced via the two gluons
diagram with strong vertices (see, for example, the diagram given
by {\bf Fig.~10}.)

\bfi \centering
\includegraphics[height=70mm,keepaspectratio=true,angle=0]{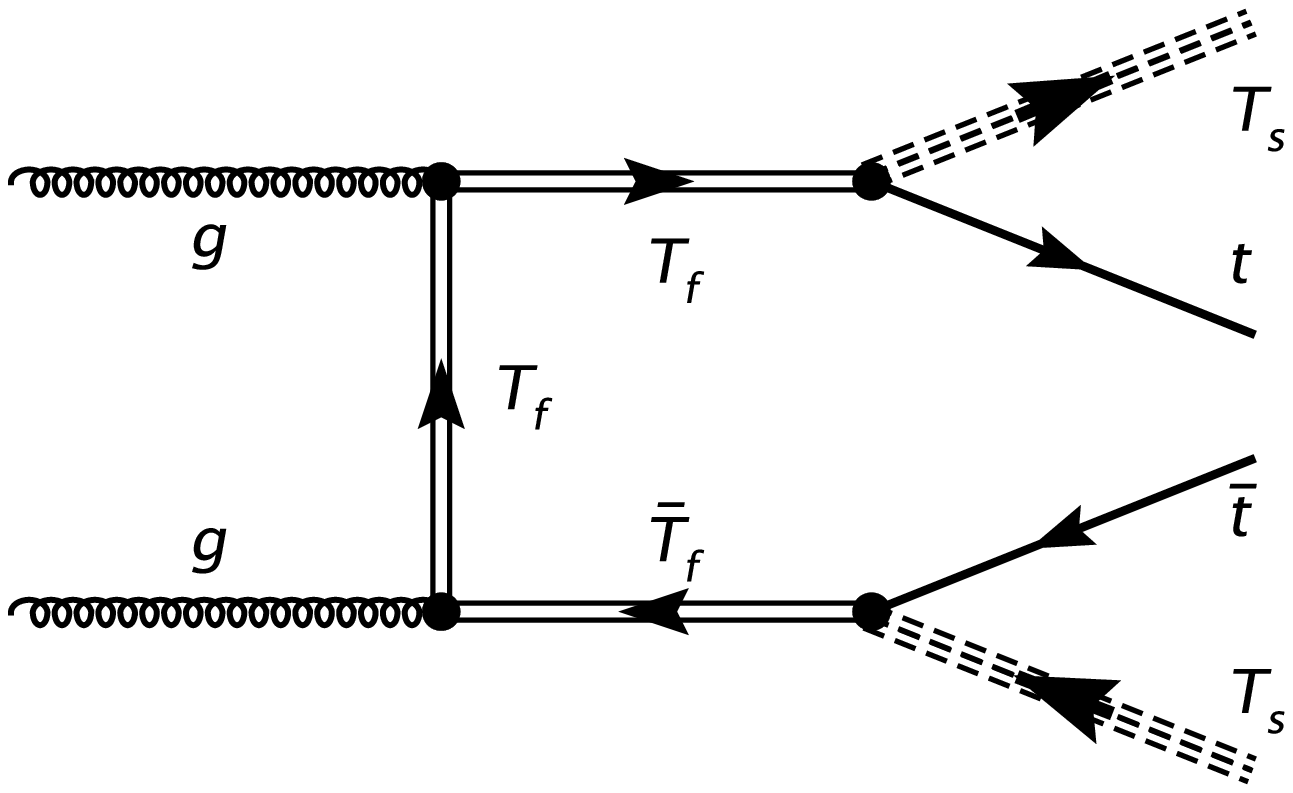}\caption{Two gluon production of T-balls.} \lb{Fig.8}\efi


Let us stress that if our description works, then the fermionic\\
``b-replaced NBS''  would make a perfect simulation of a fourth
family $\bf t'$.

It only deviates:

1) by needing either a more complicated diagrams for its
production,

2) or by the emission of soft b-quarks,

3) or by simultaneously emitted a $\bf W$-boson and $\bf T_S$:
$$ \bf ''b-replaced''\,\,\, NBS \to T_S + W.$$
Now we take in our model the particle simulating the $\bf t'$: a
fermionic ``b-replaced NBS'', for instance,  $$ \bf
T_f(b-replaced) = 5t + b + 5\bar t,$$ or $$ \bf T_f(b-replaced,
n_ab\bar b) = 5t + b + n_a b\bar b + 5\bar t,\quad {\mbox{etc}}.$$
So then we could really claim: {\large \bf we expect that the
Tevatron-LHC experiments should find or a fourth family t'-quark,
or the fermionic ``b-replaced NBS'', or both of them.}

We have shown that the bound states $\bf T_S(b-replaced)$ and $
\bf T_S(b-replaced, n_ab\bar b)$ can have masses very close to
''11'' NBS: $\bf M_{T_f}$ ($\sim 300-400\,\,{\mbox {GeV}}$, by our
prediction). We also have considered more heavy ''b-replaced'' NBS
(T-balls) with $\bf M_T > 400\,\,{\mbox {GeV}}$. All of them can
be investigated at LHC.

Concerning how to distinguish the two hypotheses:

I) the true fourth family,

II) our bound state ``b-replaced NBS'',

we can immediately say: we do not expect exactly the same
cross-section times branching ratio as that to be estimated for
the true fourth family. Thus if the cross-section agrees extremely
precisely with the calculation for a simple fundamental fourth
family t'-quark, then it is suggested that our model is not the
right explanation, but if it is in the same range, but do not
match perfectly, then it would support our model.

There are several deviations in the case of the ``b-replaced NBS''
particle production:

A) form-factors;

B) the soft b-emission;

C) some diagrams not having analogues in the fundamental $\bf t'$
production;

D) possibly alternative decays.

\section{\un{The Tevatron-LHC experiments searching for W, Z, t, t'} \un {and different
jets.}}

This talk, as the talk by Holger Bech Nielsen at CERN \ct{1}, is
devoted to the main purpose of the experiment -- to search for the
Higgs boson, and in this connection to search for T-balls, just
what we were considering as the b-replaced NBSs.

From the beginning, we have considered the following NBSs:
$$\bf 6t + 6\bar t, \qquad  6t + 5\bar t.$$
First of these NBS is a scalar boson $\bf T_S$, and the second one
is a fermion $\bf T_f$ with quantum numbers of t-quark, which is
difficult to distinguish from the quark of the fourth generation.

A typical process which is observed at the Tevatron ($\bf p\bar
p$-collisions, $\bf \sqrt s \backsimeq 1.96$ GeV) is shown in {\bf
Fig.~11}. Unfortunately, the cross-section for the Higgs boson
production at the Tevatron is predicted to be rather small and
sufficient data for a discovery of H is unlikely to be collected
before the date when more powerful LHC experiment begins to work
in 2008.


\bfi \centering
\includegraphics[height=120mm,keepaspectratio=true,angle=0]{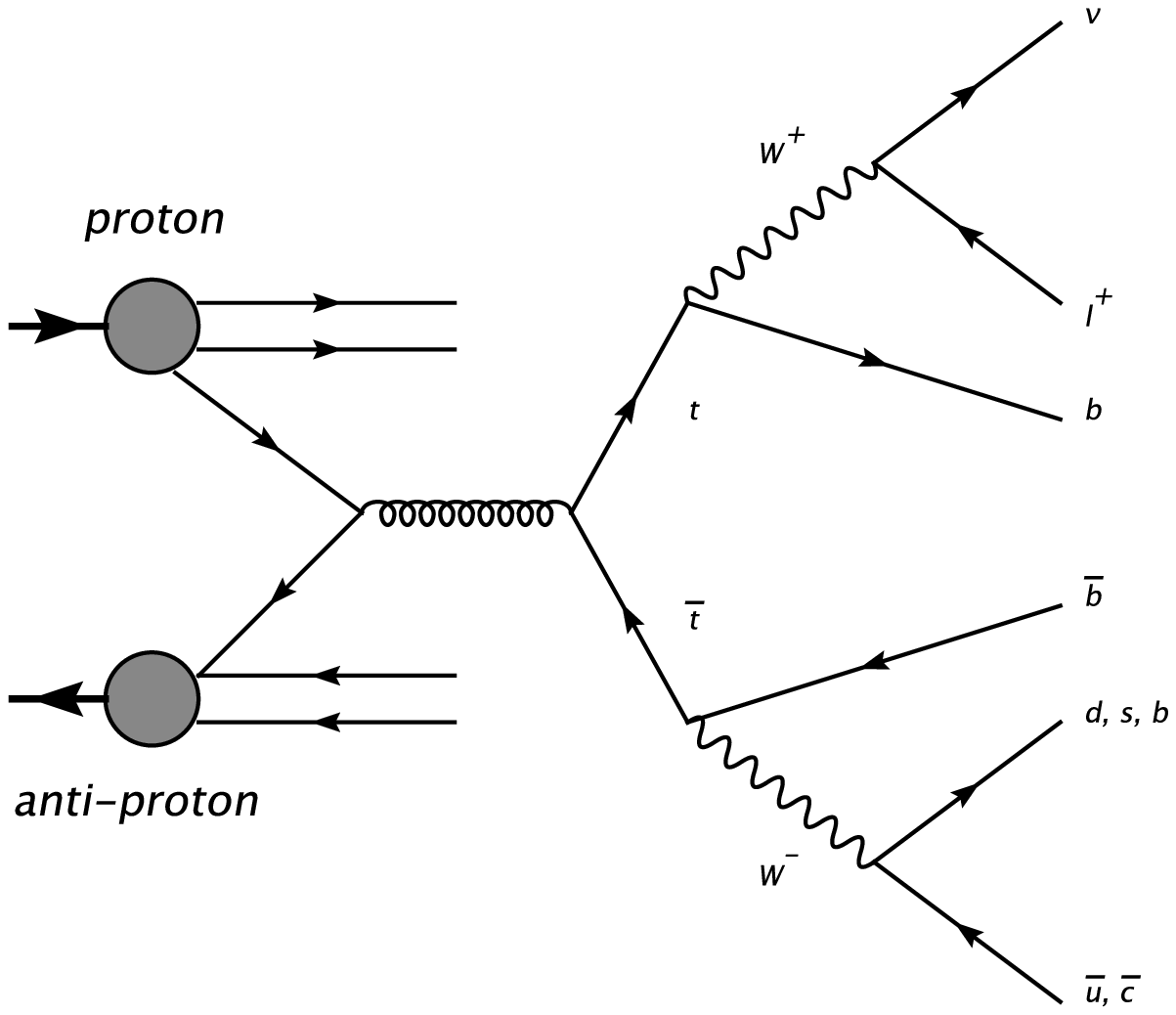}\caption{}\efi

\clearpage \newpage

There are several Higgs production methods at the LHC, which lead
to the observable Higgs production cross-sections $\bf \sigma
(pp\to HX).$ These include:

$\bullet$ gluon-gluon fusion;

$\bullet$ WW and ZZ fusion;

$\bullet$ Associated production of W and Z bosons;

$\bullet$ Associated production of $\bf t\bar t$, or $\bf t'\bar
t'$.

Typical Feynman diagrams for the signal and background processes
are shown in {\bf Figs.~12 and 13}.

\bfi \centering
\includegraphics[height=120mm,keepaspectratio=true,angle=0]{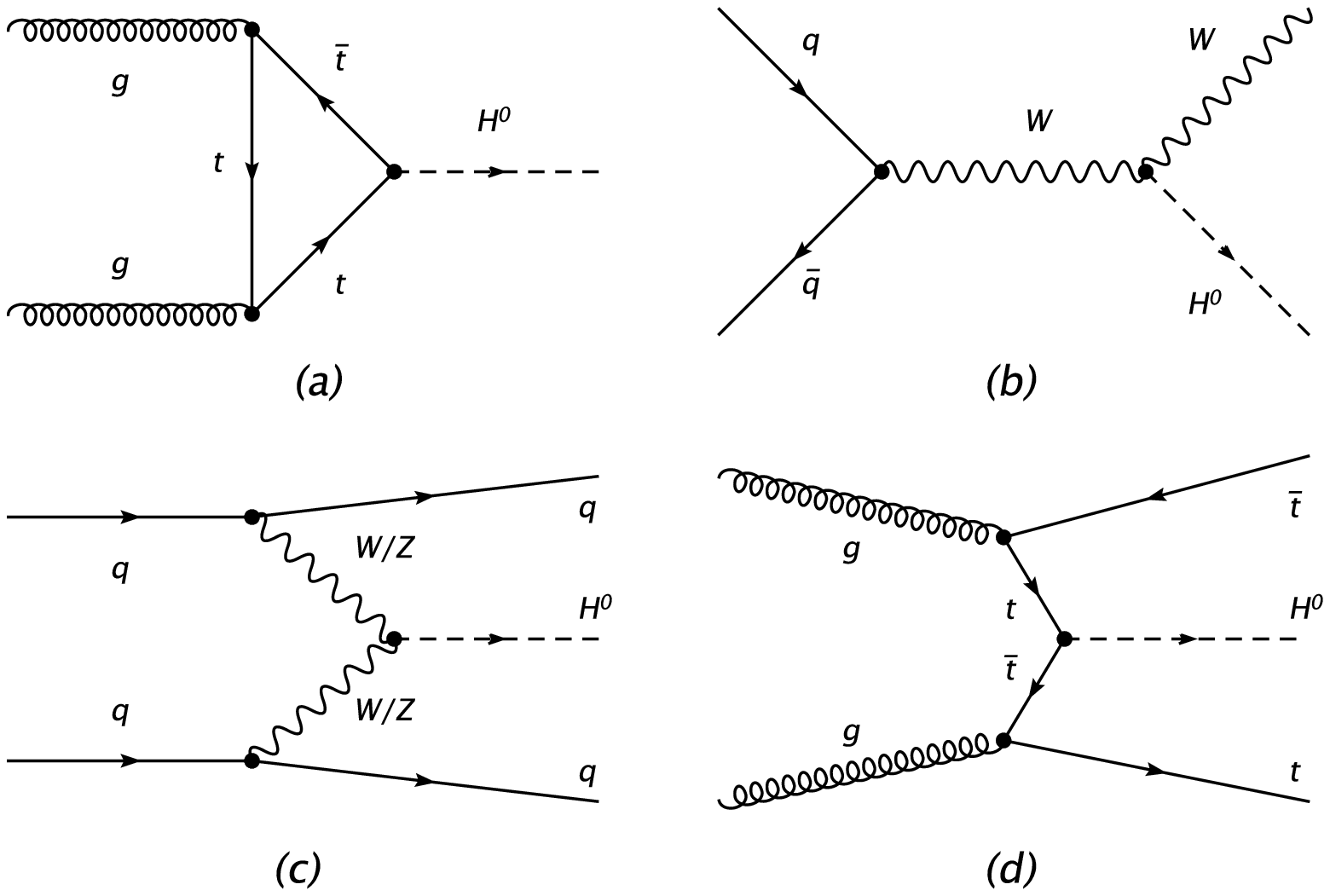}\caption{Feynman
diagrams for the processes (a) $gg\to t\bar tH^0$, (b) $q\bar q
\to t\bar tH^0$, (c) $gg\to t\bar tb \bar b$, (d) $gg\to
Z/W/\gamma^* \to H^0(t\bar t\,\,\, or\,\,\, b \bar b)$.}\efi

\bfi \centering
\includegraphics[height=120mm,keepaspectratio=true,angle=0]{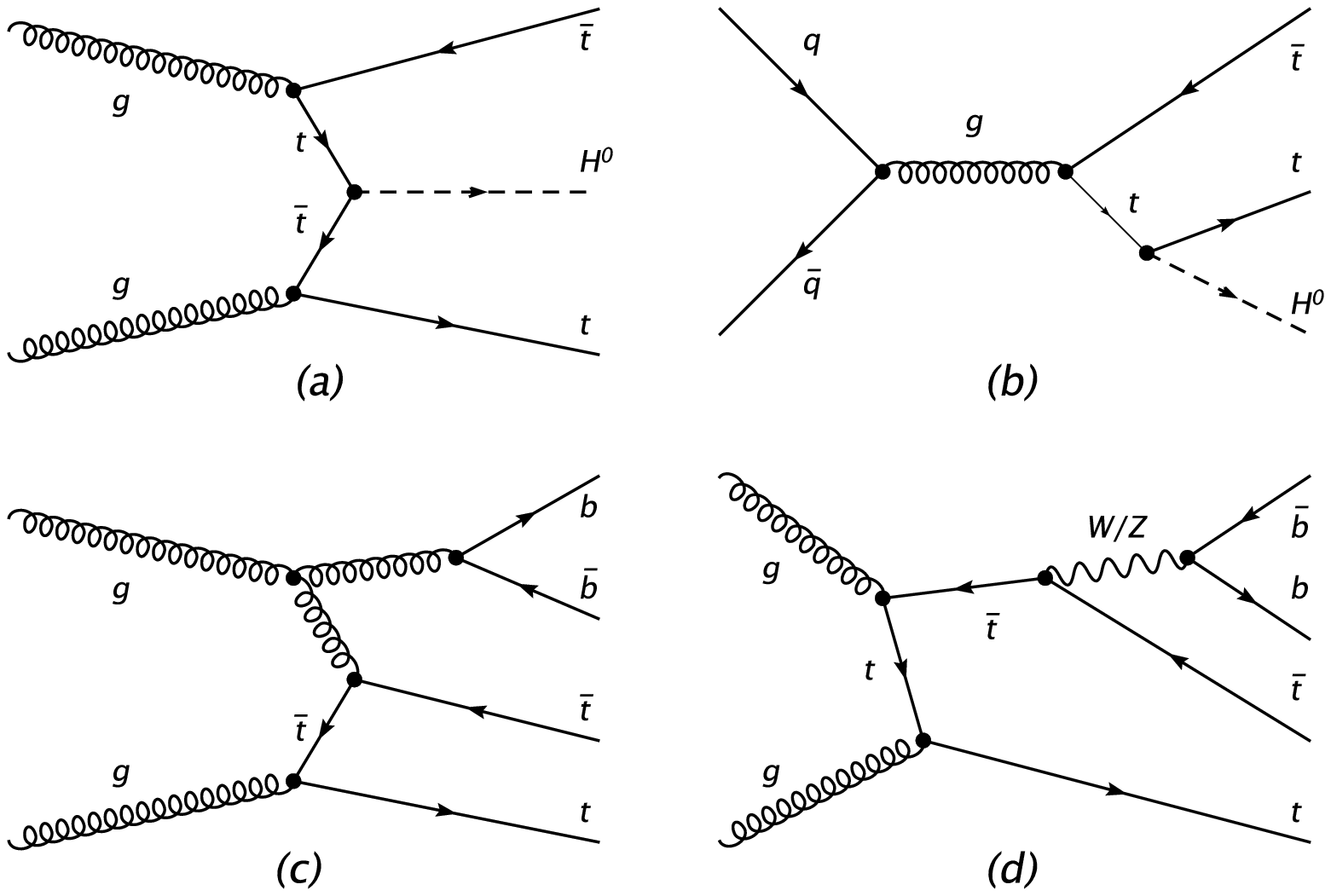}\caption{Feynman
diagrams for the production of leptons and jets.}\efi

\clearpage \newpage

At the LHC, $t\bar tH^0$ is produced 90\% of the time via a
gluon-gluon interaction and only by a qq-interaction in the
remaining 10\%.

{\large \bf Once produced, a top-quark decays almost exclusively
to the W-boson and b-quark.}

W-bosons decay hadronically about 2/3 of the time, producing two
jets in the final state.

The branching ratios for these processes are shown \\ in the{\bf
Table 1:}

\bc {\large \bf Table 1}\ec
$$ \bf  t\to Wb \qquad 0.998,$$
$$ \bf  W\to l\nu \qquad 0.108,$$
$$ \bf  W\to hadrons \qquad 0.676,$$
$$ \bf  t\bar t\to l\nu bjj\bar b \qquad 0.291.$$
The final state with the highest branching fraction is where both
top-quarks decay hadronically, producing light-jets and two
b-jets. When the decay of the Higgs boson to the two b-quarks is
taken into account, this produces a purely hadronic final state.
Requiring one of the W-bosons to decay leptonically produces a
final state with four b-jets, two light-jets, one lepton missing
momentum (see {\bf Fig.~14}).

Only $\bf l$ and $\bf \mu $ ($l=e, \,\,\mu$) are considered in
this analysis.

\bfi \centering
\includegraphics[height=120mm,keepaspectratio=true,angle=0]{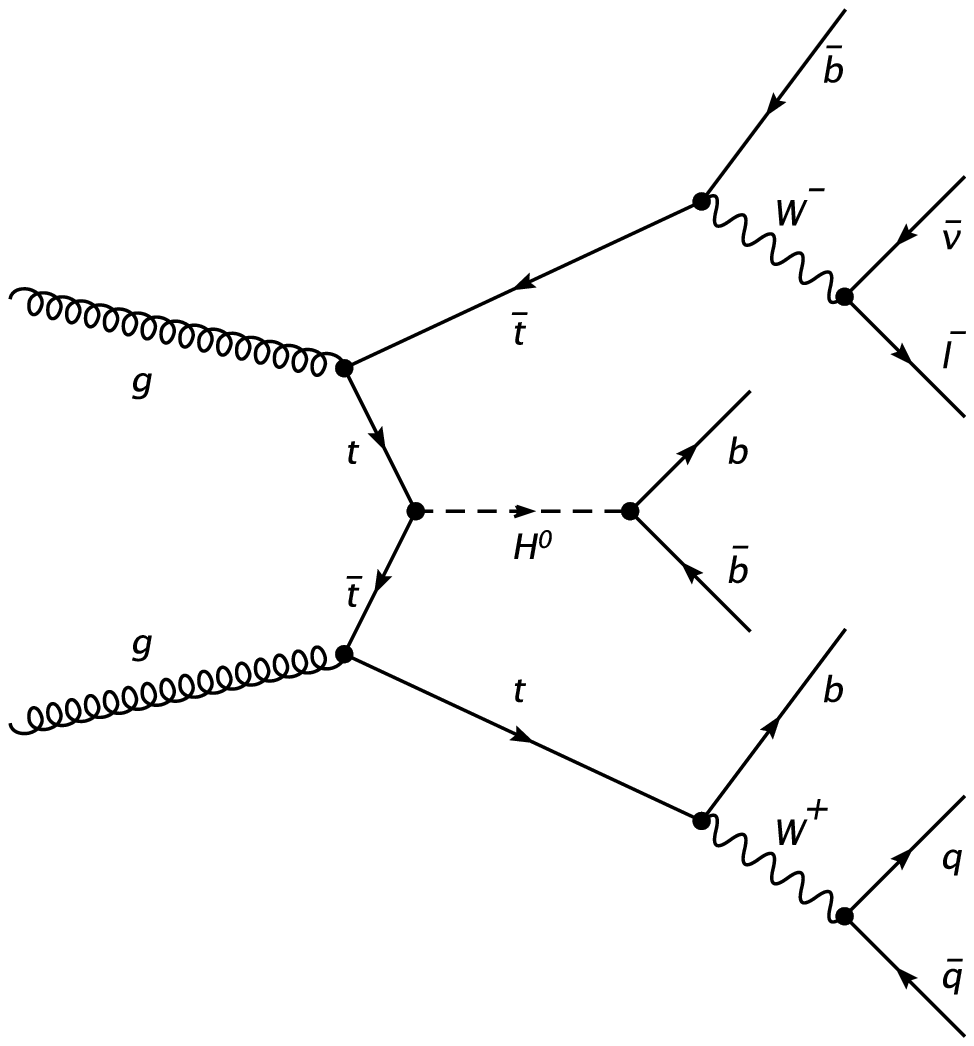}\caption{Feynman
diagrams for the production of leptons and jets.}\efi

\section{\un{Can we see T-balls at LHC or Tevatron?}}

At present, the first question is: can we observe the NBSs T-balls
at LHC or Tevatron?

If the mean square radius of the T-ball is small in comparison
with its Compton wave length: \be \bf r_0\approx (\sqrt 2
M_t)^{-1} << \frac{1}{m_{NBS}},\lb{r0} \ee then the NBS can be
considered as an almost fundamental particle.  Then our NBS are
strongly bound and can be observed at colliders. As $\bf t'$-quark
of the fourth generation the fermionic NBS $\bf T_f$ will belong
to the fundamental representation {\un 3} (color triplet).

What processes with the participation of T-balls have to play the
main role in the experiments at colliders?

A) First, the possible decay mechanism: \be \bf  H\to 2T_S,
\lb{dH} \ee if $ \bf  M_{T_S} < \frac 12 m_H.$

Using limits given by the Tevatron experiments: \be \large \bf 120
\lesssim m_H \lesssim 160\,\,\, GeV,  \lb{mH} \ee we obtain the
requirement for the Higgs decay mechanism to work (on shell):
$$\bf  M_{T_S} \lesssim 80 \,\,\, GeV.$$

If $\bf  M_{T_S} > \frac 12 m_H$, then the decay (\ref{dH}) is
absent in nature, and in the above-mentioned process the $\bf
T_S$-balls fly away forming jets. As a result, we have the
production of hadrons with high multiplicity :
$$ \bf T_S(b-replaced) \to JETS.$$
Jets create a lot of hadrons \footnote{In Ref.~\ct{8} Li and one
of us (H.B.N) have though argued that for the very light bound
states the number of jets will be more moderate and the number of
hadrons not so enormous again.} .

Since the coupling of H with T-balls is very strong, then {\large
\bf the total decay width of the Higgs boson is enlarged due to
the decays of H
 in $\large \bf T_S$,}
while the decay width of H into leptons and photons (these
channels are easily observed experimentally) is essentially less.

We see now that the present theory of T-balls predicts the
observation of the Salam-Weinberg Higgs boson H as a broad peak at
colliders.

B) Second,  the all processes with the replacement $\bf t\bar t
\to T_f\ov{T_f}$ (see, for example, {\bf Fig.~10}.)

In the most optimistic cases the NBS $\bf 6t + 5\bar t$ plays a
role of the fundamental quark of the fourth generation, say, with
mass $\simeq 300$ GeV, given by our preliminary estimate.

The most important production mechanism for producing pairs of
T-balls is \\ {\large \bf two initial gluons} that must be
provided, say, from the Tevatron hadronic  $\bf p + \bar p$
collisions (see Refs.~\ct{8} and {\bf Fig.~10}).

According to the diagram of {\bf Fig.~10} , the following decay is
possible to observe at high energy colliders:
$$\bf T_f(b-replaced) \to T_S + t + n_a b\bar b,$$
since the mass of $\bf T_S$ is expected to be less than the $\bf
T_f$-mass.

\clearpage \newpage

\section{\un{CDF II Detector experiment searching for}\\\un{heavy
top-like quarks at the Tevatron.}}

Have we seen at colliders  $\bf t'$ or $ \bf T_f$, or both of
them?

Recent experiments with CDF II Detector of the Tevatron \ct{21,22}
do not exclude the existence of $\bf t'$ or T-balls if the mass is
over $284\,\,{\mbox{GeV}}$ (see {\bf Figs.~15-18}). Here we shall
argue for that the very strange events, observed at the Tevatron
as a fourth family $\bf t'$, which decays into a $\bf W$ and a
presumed quark-jet, might in our model find another explanation:
maybe it is the decay of a b-replaced NBS (T-ball) into a $ \bf W$
and a gluon jet. But, of course, it is very difficult to measure
whether the jet coming out is from a gluon or quark.


\bfi \centering
\includegraphics[height=120mm,keepaspectratio=true,angle=0]{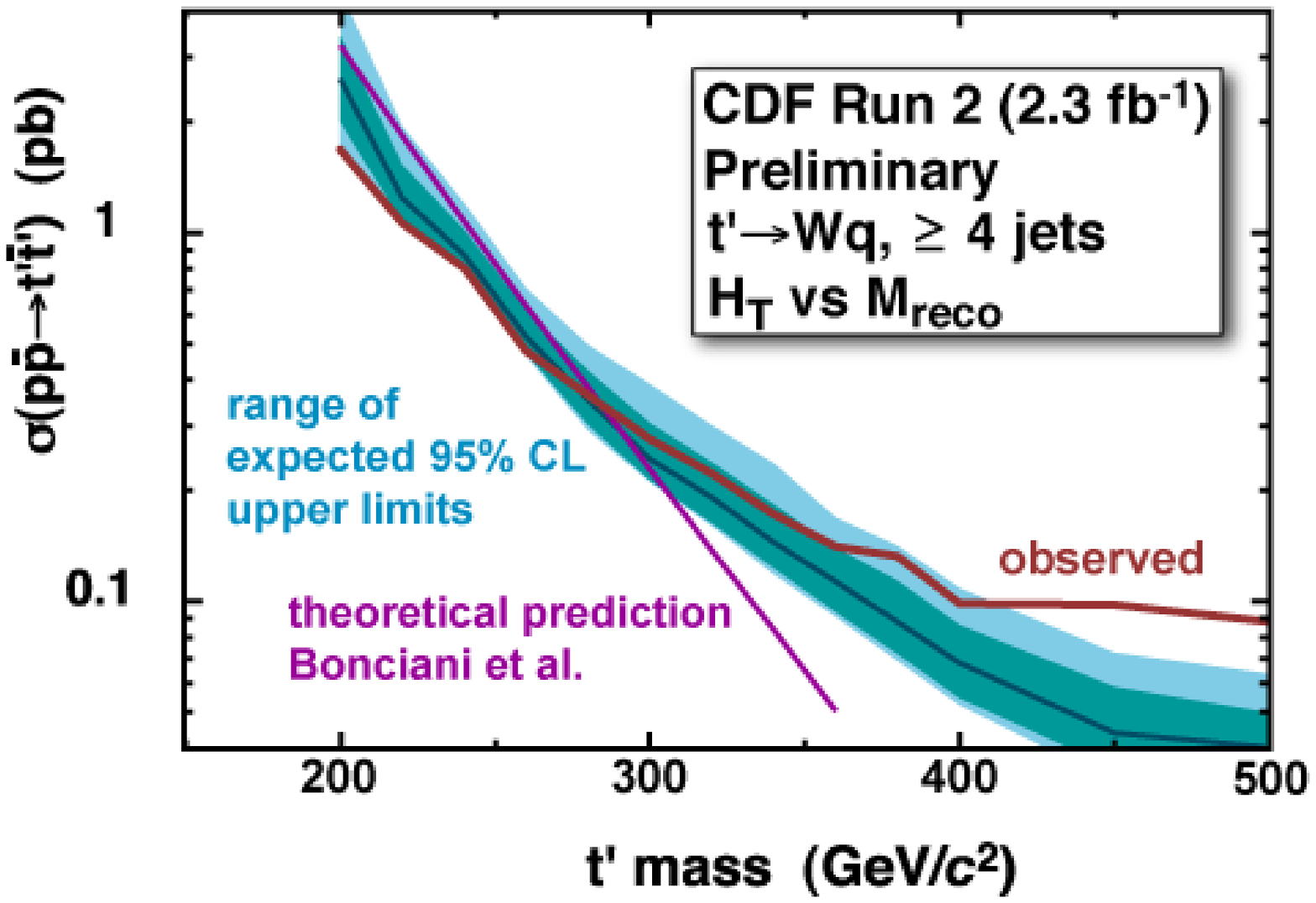}
\caption{Upper limit, at 95\% CL, a fourth-generation t' quark
with a mass below 284 GeV is excluded.}\efi

\clearpage \newpage

\bfi \centering
\includegraphics[height=120mm,keepaspectratio=true,angle=0]{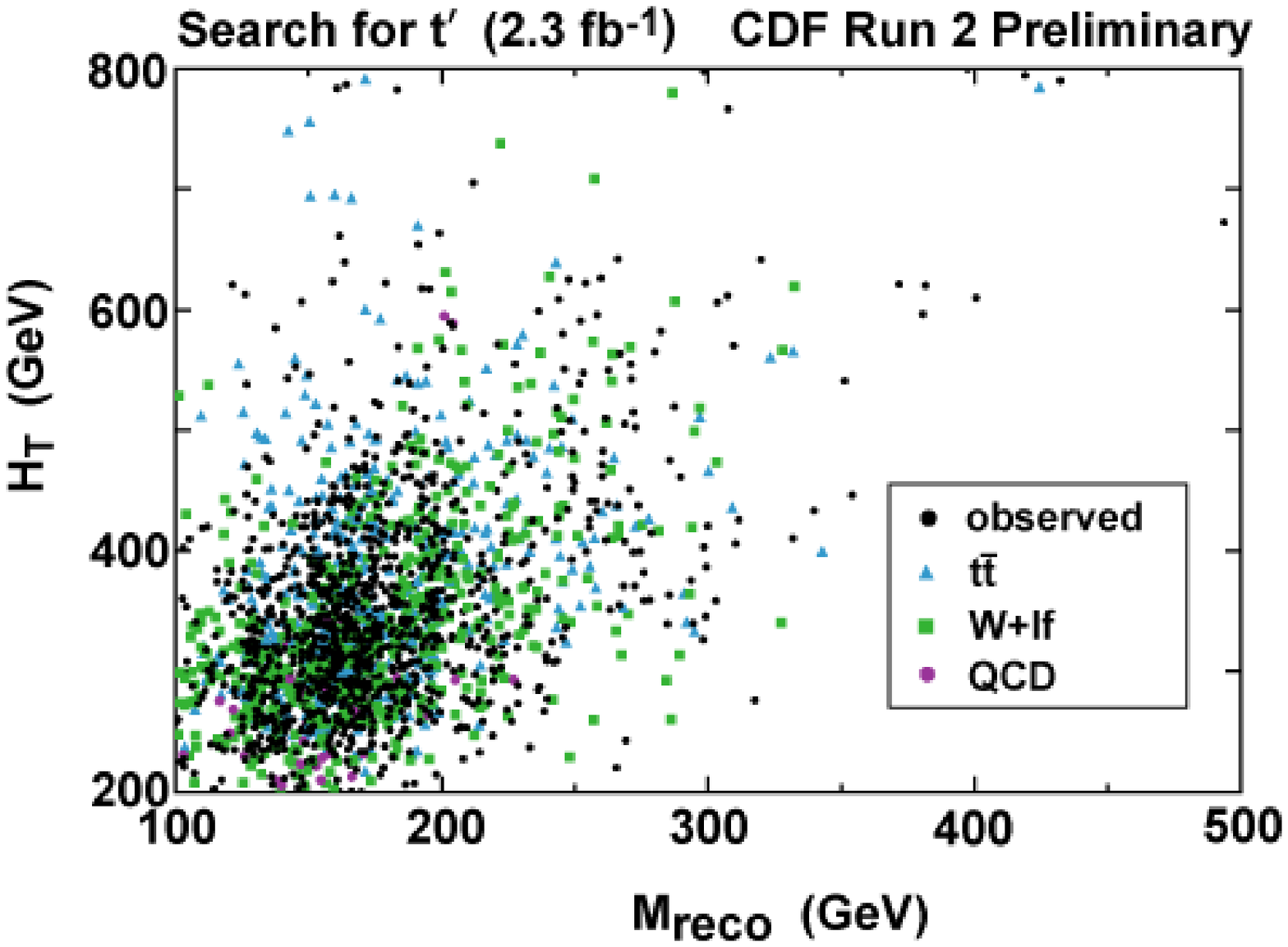}
\caption{2D distribution of $H_T$ vs $M_{rec}$ distribution
showing the data (black points) and the fitted number of
background events; QCD (purple circles), W+JETS (green squares)
and $t'\bar t'$ (blue triangles)}\efi

\clearpage \newpage

\bfi \centering
\includegraphics[height=120mm,keepaspectratio=true,angle=0]{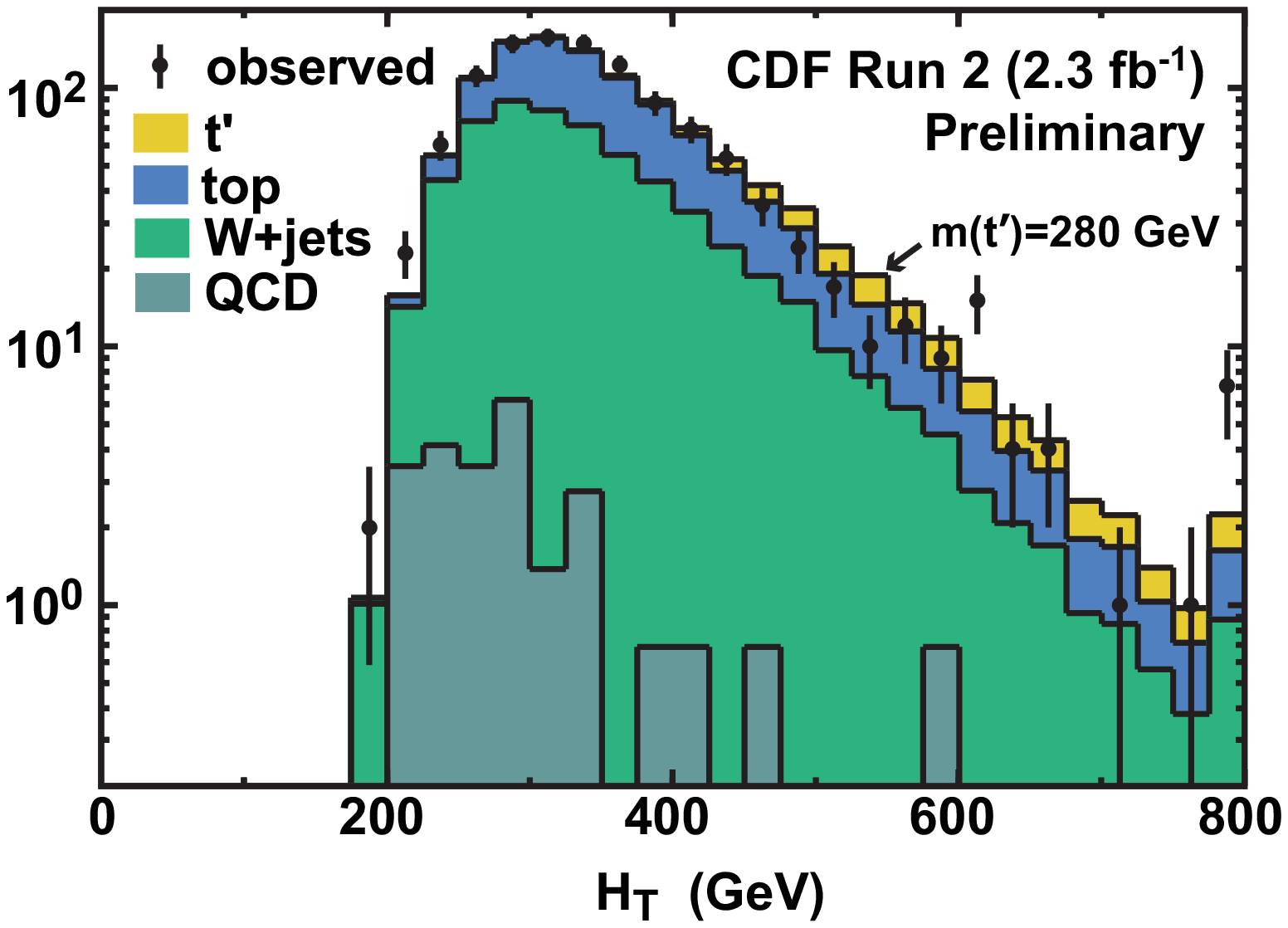}
\caption{}\efi

\bfi \centering
\includegraphics[height=120mm,keepaspectratio=true,angle=0]{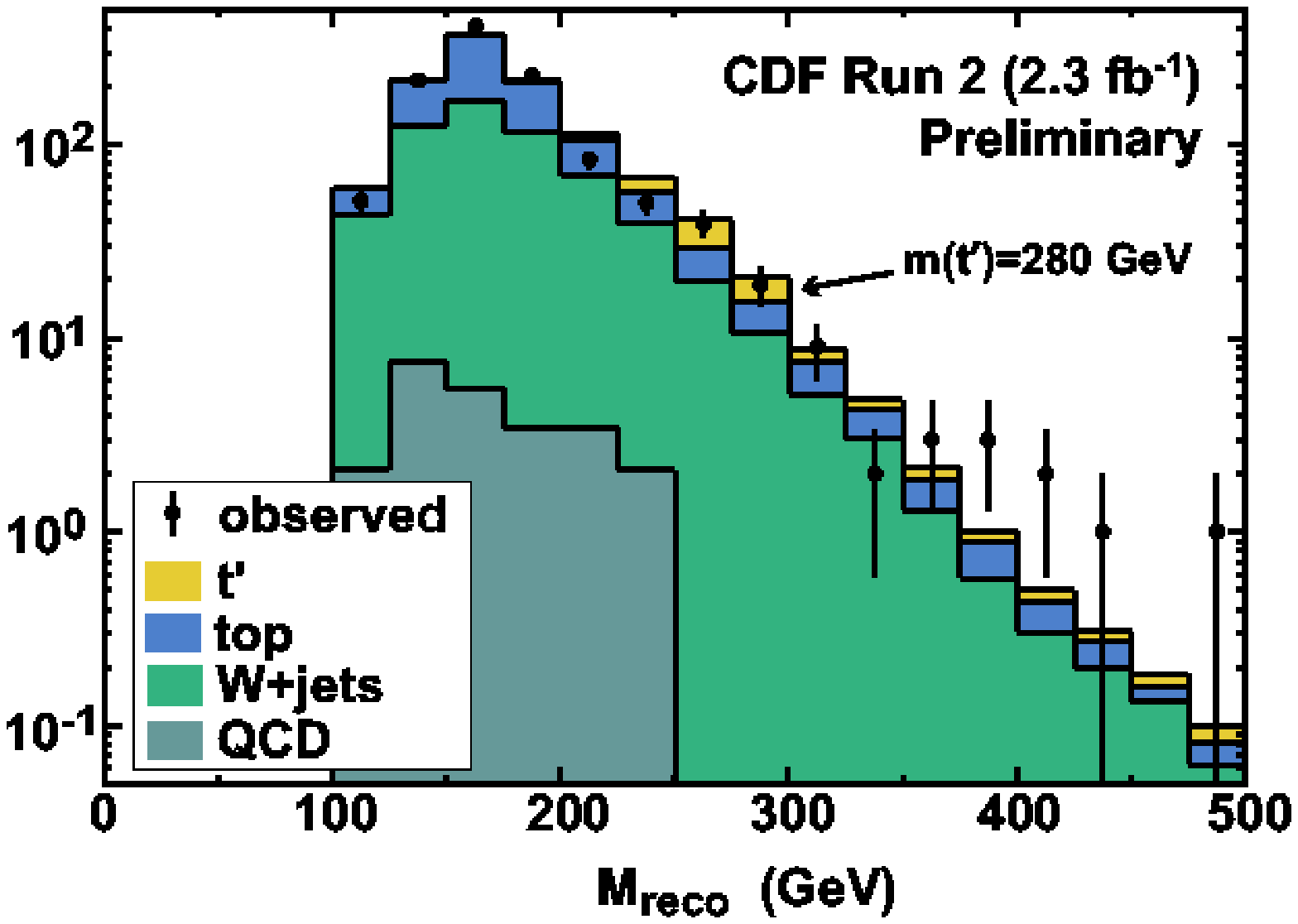}
\caption{}\efi

\clearpage \newpage

\section{\un{Charge multiplicity in decays of T-balls.}}

Actually Li and Nielsen suggested in Ref.~\ct{8} that the NBSs
would decay to a rather low number of jets, but at first one might
very reasonably think that since we have to do with bound states
of very many constituents and actually $\bf 6t\bar t$ pairs, it
sounds that the possibility of them decaying into as many jets as
there are pairs to annihilate, say - or even the number of
constituents - has some intuitive appeal and should not just be
thrown away as a possibility by the Li-Nielsen rather non-safe
argument. We shall therefore here develop what we would expect in
the case of the separate $\bf t\bar t$ pairs decaying essentially
separately, although we do not really believe that any longer: if
the mass of the NBS, containing 6 pairs of $\bf t\bar t$, is $\bf
M_S$, then the energy per one annihilation of $\bf t\bar t$
approximately is equal to the following value: \be  \bf E_{an}=
E_{(for\,\, one\,\, annihilation)}\approx \frac 16 M_S, \lb{15}
\ee e.g.
$$ \bf E_{(for\,\, one\,\, annihilation)}\approx 10\,\,
GeV,$$ if $$ \bf M_S\approx 60\,\,{\mbox{GeV}}.$$ In this case,
during the annihilation produced by $ \bf e^+e^-$-collisions, the
special charge multiplicity is
$$\bf  <N_{ch}(e^+e^-)>\approx 10,$$ while the annihilation produced
by $\bf pp$-collisions, the special charge multiplicity is
$$\bf <N_{ch}(pp)>\approx 6.$$

Such calculations of $\bf <N_{ch}>$ vs $\bf E_{an}$ are based on
the investigation of Ref.~\ct{22}. Here for $\bf M_S\approx 60\,\,
GeV$ we obtain the following values for the charge multiplicity:
\be
  \bf  N_{ch}(e^+e^-)\approx 6\cdot 10\approx 60, \lb{16} \ee

\be  \bf
         N_{ch}(pp)\approx 6\cdot 6\approx 36. \lb{17} \ee
The value of the charge multiplicity weakly depends on the NBS
mass.
For instance, if $ \bf M_S\approx 80\,\, GeV$, then:
$$ \bf <N_{ch}(pp)>\approx 6.5,$$ and
\be  \bf
         N_{ch}(pp)\approx 6\cdot 6.5\approx 39. \lb{18} \ee
But if $ \bf M_S\approx 100\,\, GeV$, then:
$$ \bf <N_{ch}(pp)>\approx 7,$$ and \be  \bf
         N_{ch}(pp)\approx 6\cdot 7\approx 42. \lb{19} \ee
However, such a maximally possible charge multiplicity will not be
realized in practice, because between the produced in the final
state pairs $\bf t\bar t$, or $ \bf b\bar b$, can exist extra
exchanges by gluons and the Higgs bosons giving new annihilations.
And we shall obtain less jets.

Indeed, it would be very strange if the decay width of the T-balls
was small. Then we would have narrow peaks in JETS. It would be
exactly a good way to see that our model were right if you could
find some narrow peak in the distribution of the total mass of
some JETS.

For $\bf pp$-collisions the estimates \ct{8} give : \be  \bf
\frac{dN_{ch}}{d\eta}|_{max}\approx 6. \lb{20} \ee Such a value is
expected for this derivative at LHC (see {\bf Fig.~19}). The
maximum of this curve corresponds to the LHC energy $\bf W=14
\,\,{\mbox{TeV}}$ in $ \bf pp$-collisions.

\bfi \centering
\includegraphics[height=100mm,keepaspectratio=true,angle=0]{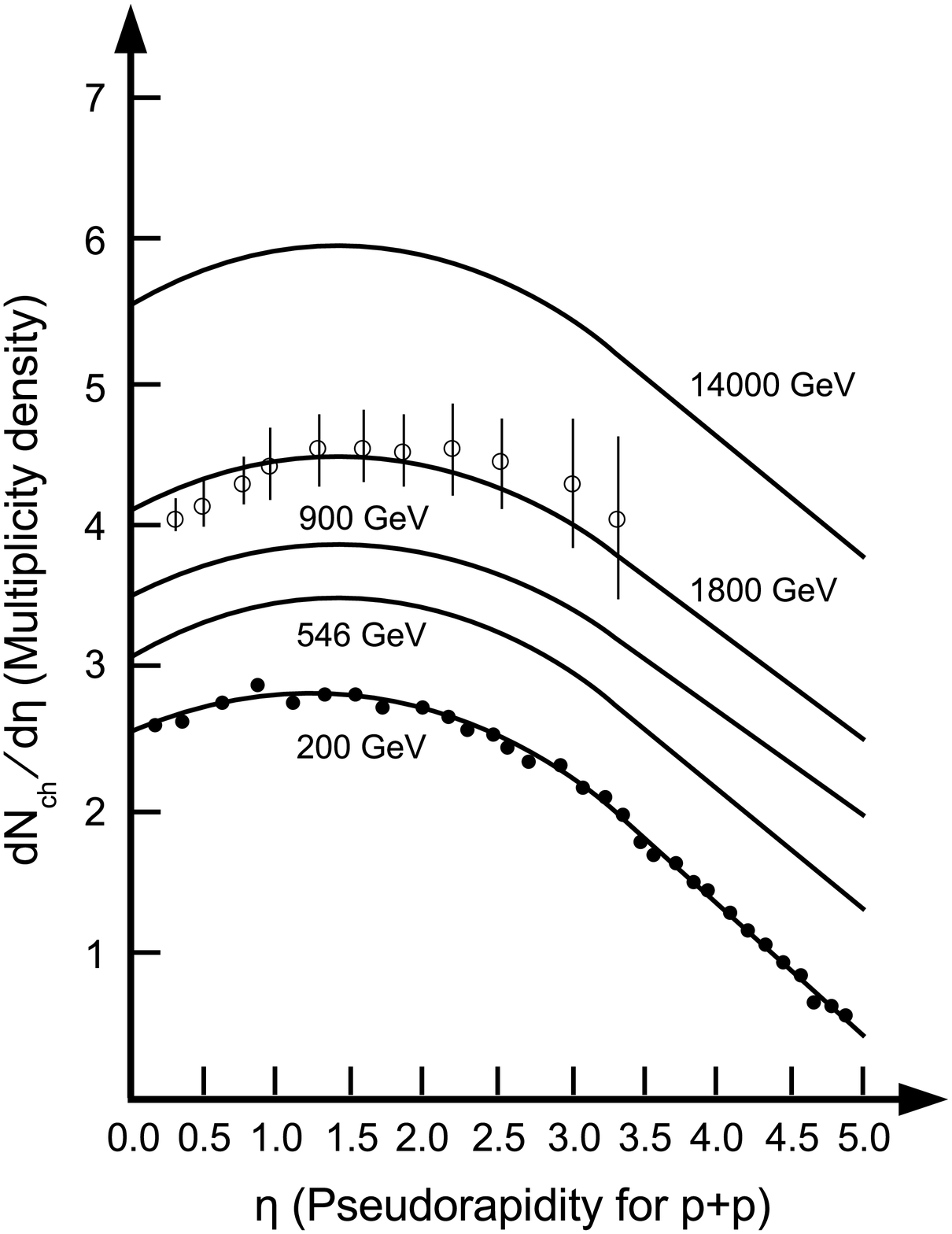}\caption{} \lb{}\efi

\clearpage \newpage

The dependence $ \bf N_{ch}$ vs $ \bf W$ is presented in {\bf
Fig.~20}. Here \be  \bf N_{ch}(pp)|_{W=14\,\, TeV}\approx 65.
\lb{21} \ee

\bfi \centering
\includegraphics[height=100mm,keepaspectratio=true,angle=0]{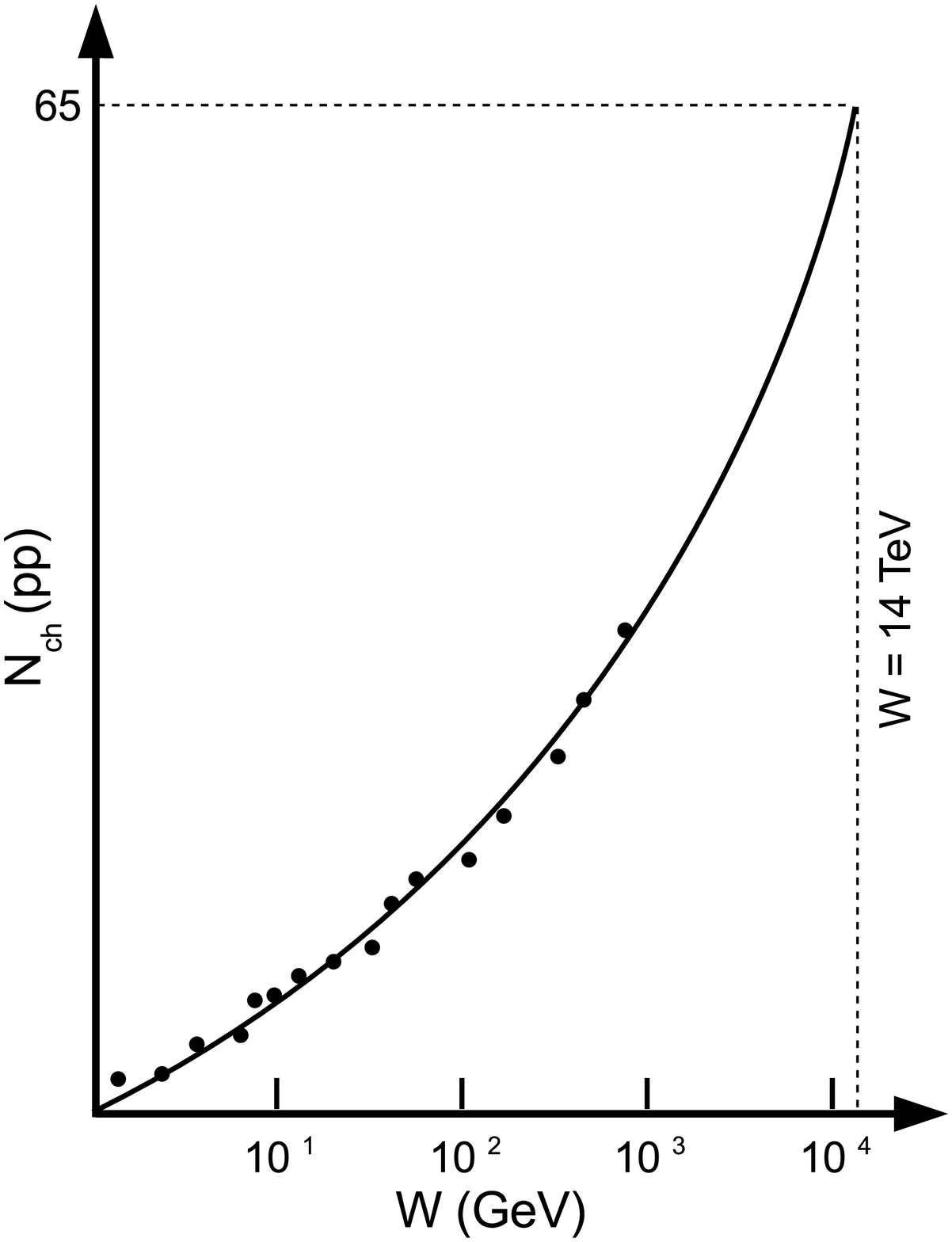}\caption{} \lb{}\efi

These calculations (figures) show that T-balls can give an
essential contributions to charge multiplicity in $ \bf
pp$-collisions, provided that their decays really go as if each
$\bf t\bar t$ pair decayed separately and not as the recent
estimate by Li-Nielsen \ct{8}.

\clearpage \newpage

\section{\un{Conclusions.}}

\begin{itemize}

\item [1.]  The present investigation is devoted to the main
problems of the Standard Model connected with searching for the
Higgs boson and based on the following three assumptions:

a) there exists $\bf 1S$--bound state of $\bf 6t+6\bar t$, e.g.
bound state of 6 quarks of the third generation with their 6
anti-quarks;

b) the forces which bind these top-quarks are so strong that they
almost completely compensate the mass of the 12 top-quarks forming
this bound state;

c) such strong forces are produced by the Higgs interactions: the
interactions of top-quarks via the virtual exchange of the scalar
Higgs bosons coupling with a large value of the top-quark Yukawa
coupling constant $\bf g_t$.

A new bound state $\bf 6t+6\bar t$, which is a color singlet, was
first suggested {\bf  by Froggatt and Nielsen} and now was named
T-ball, or T-fireball.

\item [2.] Our theory also predicts the existence of the new bound
state (NBS) $\bf 6t + 5\bar t$, which is a color triplet and a
fermion similar to the quark of the fourth generation having
quantum numbers of t-quark.

\item [3.] We have also considered ''b-replaced'' NBSs:
$$ \bf T_S(n_bb-replaced) = n_b b + (6t + 6\bar t - n_b t)$$
and $$\bf T_f(n_bb-replaced) = n_b b + (6t + 5\bar t - n_b t),$$
where $\bf n_b$ is the integer number. The presence of b-quarks in
the NBS leads to the dominance of the isospin singlets: with the
inclusion of both b and t quarks we obtain a more weak isospin
invariant picture.

\item [4.] We have estimated the masses of the lightest
''b-replaced'' NBSs:
$$\bf M_{T(b-replaced)} \simeq (300 - 400) \,\,{\mbox{GeV}},$$
and predicted the existence of the more heavy ''b-replaced'' NBSs:
$$\bf M_{T(n_bb-replaced)} > 400 \,\,{\mbox{GeV}}$$
with $\bf n_b > 1$.

\item [5.] We have developed a theory of the T-ball's condensate,
and predicted the possibility of the existence of the three SM
phases at the EW-scale, calculating the top-quark Yukawa coupling
constant at the border of the two phases (with T-ball's condensate
and without it) equal to: $ \bf g_t \approx 1$.

\item [6.] It was shown that CDF II Detector experiment searching
for heavy top-like quarks at the Tevatron ($\bf p\bar
p$-collisions, $ \bf \sqrt s \backsimeq 1.96$ GeV) can observe $
\bf T_f$-balls with masses up to 400 GeV.

\item [7.] We have considered all processes with T-balls, which
can be observed at LHC, especially the decay $$\bf H\to 2T_S$$ and
the production of $$ \bf T_f,\ov{T_f}$$ as an alternative of the $
\bf t'\ov{t'}$ production (where $ \bf t'$ is the quark of the
fourth generation with t-quark quantum numbers).

\item [8.] We have estimated the charge multiplicity at the energy
W=14 GeV at LHC: $$\bf N_{ch}(pp)|_{W=14\,\,{\mbox{TeV}}}\approx
65,$$ and have shown that the charge multiplicity coming from the
T-ball's decays is of order of this value.

\item [9.] In this investigation we have argued that T-balls can
explain why it is difficult to observe the Higgs boson H at
colliders as sharp peak: T-balls can strongly enlarge the decay
width of the Higgs particle.

\end{itemize}

\clearpage \newpage

\section{\un{Appendix. The Standard Model Lagrangian.}}

 The standard model is a gauge theory of the microscopic
interactions. The strong interaction part (QCD) is described by
the Lagrangian \begin{equation} \L_{SU_3} = - \frac{1}{4}
F^i_{\mu\nu} F^{i\mu\nu} + \sum_r \bar{q}_{r\alpha} i
\not{\!\!D}^\alpha_\beta \,q^\beta_r, \label{eq1:c20}
\end{equation}
where $g_s$ is the QCD gauge coupling constant,
\begin{equation} F^i_{\mu\nu} = \partial_\mu G^i_\nu - \partial_\nu G^i_\mu -
g_s f_{ijk}\; G_\mu^j\; G_\nu^k  \label{eq1:c21} \end{equation} is
the field strength tensor for the gluon fields $G^i_\mu, \; i = 1,
\cdots, 8$,  and the structure constants $f_{ijk}$
 $  \bf(i, j, k = 1, \cdots, 8)$ are defined by
\begin{equation} [\lambda^i, \lambda^j] = 2 i f_{ijk} \lambda^k, \end{equation}
where $\lambda^i$ are the Gell-Mann  $SU_3$ matrices.

    The $F^2$ term leads to three and four-point gluon
self-interactions. The second term in $\L_{SU_3}$ is the gauge
covariant derivative for the quarks: $q_r$ is the $r^{\rm th}$
quark flavor, $\alpha, \beta = 1,2,3$ are color indices, and
\begin{equation} D^\alpha_{\mu \beta} = (D_\mu)_{\alpha \beta} =
\partial_\mu \delta_{\alpha \beta} + i g_s \;G^i_\mu\; L^i_{\alpha\beta},
\label{eq1:c22} \end{equation} where the quarks transform
according to the triplet representation matrices $L^i
={\lambda^i}/{2}$. The color interactions are diagonal in the
flavor indices, but in general change the quark colors. They are
purely vector (parity conserving). There are no bare mass terms
for the quarks in (\ref{eq1:c20}). These would be allowed by QCD
alone, but are forbidden by the chiral symmetry of the electroweak
part of the theory.  The quark masses will be generated later by
spontaneous symmetry breaking.  There are in addition effective
ghost and gauge-fixing terms which enter into the quantization of
both the $SU_3$ and electroweak Lagrangians, and there is the
possibility of adding an (unwanted) term which violates $CP$
invariance.

The electroweak theory is based on the $SU_2 \times U_1$
Lagrangian:
\begin{equation} \L_{SU_2 \times U_1} = \L_{\rm gauge} + \L_\varphi + \L_f + \L_{\rm
Yukawa}. \label{eqch10b} \end{equation} The gauge part is
\begin{equation} \L_{\rm gauge} = - \frac{1}{4} F^i_{\mu \nu} F^{\mu \nu i} -
\frac{1}{4} B_{\mu \nu} B^{\mu \nu}, \label{eqch11b}
\end{equation} where $W^i_\mu, \; (i = 1, \; 2, \; 3)$ and $B_\mu$
are respectively the $SU_2$ and $U_1$ gauge fields, with field
strength tensors
\begin{eqnarray} B_{\mu \nu} &=& \partial_\mu B_\nu - \partial_\nu
B_\mu \nonumber \\
F_{\mu \nu} &=& \partial_\mu W_\nu^i - \partial_\nu W_\mu^i - g
\epsilon_{ijk} W^j_\mu W^k_\nu, \label{eqch12a} \end{eqnarray}
where $g (g')$ is the $SU_2$ $(U_1)$ gauge coupling and
$\epsilon_{ijk}$ is the totally antisymmetric symbol. The $SU_2$
fields have three and four-point self-interactions.

The field $B$ belongs to the $U_1$ theory and is associated with
the weak hypercharge $ Y = Q - T_3$, where $Q$ and $T_3$ are
respectively the electric charge operator and the third component
of weak $SU_2$. It has no self-interactions. The $B$ and $W_3$
fields will eventually mix to form the photon and $Z$~boson.

The scalar part of the Lagrangian is
\begin{equation} \L_\varphi = (D^\mu \varphi)^{\dag} D_\mu \varphi - V(\varphi),
\label{eqch13b} \end{equation} where $\varphi =
\left(\begin{array}{c} \varphi^+ \\ \varphi^0 \end{array} \right)$
is a complex Higgs scalar, which is a doublet under $SU_2$ with
$U_1$ charge $Y_\varphi = + \frac{1}{2} $.  The gauge covariant
derivative is
\begin{equation} D_\mu \varphi = \left( \partial_\mu + i g \frac{\tau^i}{2}
W_\mu^i + \frac{i g'}{2} B_\mu \right) \varphi, \label{eqch14b}
\end{equation} where the $\tau^i$ are the Pauli matrices.
   The square of the covariant derivative leads to three and four-point
interactions between the gauge and scalar fields.

$V(\varphi)$ is the Higgs potential.  The combination of $SU_2
\times U_1$ invariance and renormalizability restricts $V$ to the
form
\begin{equation} V(\varphi) = + \mu^2 \varphi^{\dag}\varphi + \lambda (\varphi^{\dag}
\varphi)^2. \label{eq15} \end{equation} For $\mu^2 < 0$ there will
be spontaneous symmetry breaking.  The $\lambda$ term describes a
quartic  self-interaction between the scalar fields. Vacuum
stability requires $\lambda > 0$.

The fermion term is
\begin{equation} \L_F = \sum^F_{m = 1} \left( \bar{q}^0_{mL} i
\not\!\!D q^0_{mL} + \bar{l}^0_{mL} i \not\!\!D l^0_{mL}
 + \bar{u}^0_{mR} i \not\!\!D u^0_{mR} +
\bar{d}^0_{mR} i \not\!\!D d^0_{mR} + \bar{e}^0_{mR} i \not\!\!D
e^0_{mR} \right). \label{eqch16} \end{equation} In (\ref{eqch16})
$m$ is the family index, $F \ge 3$ is the number of families, and
$L(R)$ refer to the left (right) chiral projections $\psi_{L(R)}
\equiv (1 \mp \gamma_5) \psi/2$. The left-handed quarks and
leptons
\begin{equation}
q^0_{mL}= \left( \begin{array}{c} u^0_m \\ d^0_m \end{array}
\right)_L \ \ \ \ \ l^0_{mL} = \left( \begin{array}{c} \nu^0_m \\
e^{-0}_m
\end{array} \right)_L
\end{equation}
transform as $SU_2$ doublets, while the right-handed fields
$u^0_{mR}, \; d^0_{mR}$, and  $e^{-0}_{mR}$ are singlets. Their
$U_1$ charges are  $Y_{q_L} = \frac{1}{6}, \; Y_{l_L}
=-\frac{1}{2}, \; Y_{\psi_R} = q_\psi$. The superscript $0$ refers
to the weak eigenstates, i.e., fields transforming according to
definite $SU_2$ representations.  They may be mixtures of mass
eigenstates (flavors). The quark color indices $\alpha = r, \; g,
\; b$ have been suppressed. The gauge covariant derivatives are

\begin{equation}  \begin{array}{lclclcl}
D_\mu q^0_{mL} &=& \left( \partial_\mu + \frac{i g}{2} \tau^i
W^i_\mu + i\frac{ g'}{6} B_\mu \right) q^0_{mL} &  \ \ \ \ \ &
D_\mu u^0_{mR} &=& \left( \partial_\mu + i\frac{2}{3}g' B_\mu
\right)
u^0_{mR}  \\
D_\mu l^0_{mL} &=& \left( \partial_\mu + \frac{i g}{2} \tau^i
W^i_\mu - i\frac{ g'}{2} B_\mu \right) l^0_{mL} &  \ \ \ \ \ &
D_\mu d^0_{mR} &=& \left(\partial_\mu -i\frac{g'}{3}B_\mu  \right)
d^0_{mR}  \\
& &  &  \ \ \ \ \ & D_\mu e^0_{mR} &=& \left( \partial_\mu - i g'
B_\mu     \right) e^0_{mR},
\end{array}  \label{eqch17} \end{equation}
from which one can read off the gauge interactions between the $W$
and $B$ and the fermion fields.
 The different transformations of the $L$ and $R$ fields (i.e.,  the symmetry
is chiral) is the origin of parity violation in the electroweak
sector. The chiral symmetry also forbids any bare mass terms for
the fermions.

The last term in (\ref{eqch10b}) is
\begin{equation} L_{\rm Yukawa} = - \sum^F_{m,n =1} \left[
Y^u_{mn} \bar{q}^0_{mL} \tilde{\varphi} u^0_{mR} + Y^d_{mn}
\bar{q}^0_{mL} \varphi d^0_{nR} + Y^e_{mn} \bar{l}^0_{mn} \varphi
e^0_{nR} \right] + {\rm H.C.}, \label{eqch18}
\end{equation} where the matrices $Y_{mn}$ describe the
Yukawa couplings between the single Higgs doublet, $\varphi$, and
the various flavors $m$ and $n$ of quarks and leptons.  One needs
representations of Higgs fields with hypercharges
 $Y = \frac{1}{2}$ and $-\frac{1}{2}$ to
give masses to the down quarks, the electrons, and the up quarks.
The representation $\varphi^{\dag}$ has $Y = - \frac{1}{2}$, and
$\tilde{\varphi} \equiv i \tau^2 \varphi^{\dag} = \left(
\begin{array}{c} \varphi^{0^{\dag}} \\ - \varphi^- \end{array}
\right)$ has $Y=-\frac{1}{2}$. All of the masses can therefore be
generated with a single Higgs doublet if one makes use of both
$\varphi$ and $\tilde{\varphi}$.


\begin{thebibliography}{99}

\bibitem{1} \textcolor{Red} {C.D. Froggatt, L.V. Laperashvili, R.B.Nevzorov, H.B.
Nielsen. {\it The Production of $6t+6\bar t$ bound state at
colliders.} A talk given  by Holger Bech Nielsen at CERN, 2008;
preprint CERN-PH-TH/2008-051.}
\bibitem{2}  \textcolor{Red}{
C.D. Froggatt and H.B. Nielsen. {\it Trying to understand
\\the Standard Model parameters,} Invited talk by H.B. Nielsen at
the ``XXXI ITEP Winter School of
Physics''\\ (February 18--26, 2003, Moscow, Russia),\\
Published in: {\it Surveys High Energy Phys.} {\bf 18}, 55--75
(2003); arXiv: hep-ph/0308144.}
\bibitem{3} \textcolor{Red}{C.D. Froggatt, H.B. Nielsen, L.V. Laperashvili. {\it Hierarchy-problem
and a bound state of 6 t and 6 anti-t,} in: Proceedings of Coral
Gables Conference on Launching of Belle Epoque in High-Energy
Physics and Cosmology
{\it (CG 2003)}, Ft. Lauderdale, Florida, 17-21 Dec 2003.\\
 Published in: Int.J.Mod.Phys. A {\bf 20}, 1268 (2005);\\ arXiv: hep-ph/0406110.}
\bibitem{4} \textcolor{Red}{C.D. Froggatt, L.V. Laperashvili, H.B. Nielsen. {\it A New bound state
6t + 6 anti-t and the fundamental-weak scale hierarchy in the
Standard Model,} in: Proceedings of 13th International Seminar on
High-Energy Physics: {\it QUARKS-2004}, Pushkinskie Gory, Russia,
24-30 May 2004; arXiv: hep-ph/0410243.}
\bibitem{5} \textcolor{Red}{
C.D. Froggatt. {\it The Hierarchy problem and an exotic bound
state.} GUTPA-04-12-02, Dec 2004; in: Proceedings of 10th
International Symposium on Particles, Strings and Cosmology, {\it
(PASCOS 04 and Pran Nath Fest)}, Boston, Massachusetts, 16-22 Aug
2004. Published in: *Boston 2004, Particles, strings and
cosmology*, pp.325-334;  arXiv: hep-ph/0412337.}
\bibitem{6} \textcolor{Red}{
C.D.~Froggatt, L.V.~Laperashvili, H.B. ~Nielsen. {\it The
Fundamental-weak scale hierarchy in the Standard Model.} Published
in: Phys.Atom.Nucl. {\bf 69}, 67 (2006) [Yad.Fiz. {\bf 69}, 3
(2006)]; arXiv: hep-ph/0407102.}
\bibitem{7} \textcolor{Red}{C.D.~Froggatt ,
R.B.~Nevzorov, H.B. ~Nielsen. {\it Is LHC production of bound
states of 6 top and 6 anti-top possible?} in progress, to be
published in 2008.}
\bibitem{8} \textcolor{Red}{
H.B.~Nielsen, S.Y.~Li. {\it Jet Multiplicity for Hard Bound
State,} and {\it Mass Calculations for Bound States of Several
Top- and Anti-Top Quarks}, preprints CERN, to be published in
2008.}
\bibitem{9} \textcolor{Red}{
H.~Jensen (FNAL), E.~Lytken (CERN), K.~Loureiro (CERN), J.~Conway
(U.K.), R.~Erbacher (U.K.), J.~Frost (U.K.), C.~Issever (U.K.),
A.~Parker (U.K.). {\it Searching for T-balls at LHC, }, in
progress, 2008.}
\bibitem{10a}
Particle Data Group: W.-M.~Yao {\it et al.}, J.Phys. G 33, 1
(2006).
\bibitem{10}
D.L.~Bennett, H.B.~Nielsen, Int.J.Mod.Phys. A {\bf 9}, 5155
(1994); ibid., A {\bf 14}, 3313 (1999).
\bibitem{11}
D.L.Bennett, C.D.Froggatt, H.B.Nielsen, in {\it Proceedings of the
27th International Conference on High Energy Physics, Glasgow,
Scotland, 1994}, Ed. by P.Bussey and I.Knowles (IOP Publishing
Ltd, 1995), p.557; {\it Perspectives in Particle Physics '94}, Ed.
by D.Klabu\u{c}ar, I.Picek and D.Tadi\'{c} (World Scientific,
Singapore, 1995), p.255; arXiv: hep-ph/9504294;
\bibitem{12}
C.D.Froggatt and H.B.Nielsen, {\it Origin of Symmetries} (World
Sci., Singapore, 1991).
\bibitem{13}
L.V.Laperashvili, Yad.Fiz. {\bf 57}, 501 (1994) [Phys.At.Nucl.
{\bf 57}, 471 (1994)].
\bibitem{14}
C.D.Froggatt, H.B.Nielsen, Phys.Lett. {\bf B368} (1996) 96.
\bibitem{15}
C.D.~Froggatt, L.V.~Laperashvili, R.B.~Nevzorov and H.B.~Nielsen,
Phys.Atom.Nucl. {\bf 67}, 582 (2004) [Yad.Fiz. {\bf 67}, 601
(2004)]; arXiv: hep-ph/0310127.
\bibitem{16}
L.V. Laperashvili. {\it The Multiple point principle and Higgs
bosons.} Phys.Part.Nucl. {\bf 36} S38-S40 (2005), e-Print:
hep-ph/0411177.
\bibitem{17}
C.R.~Das, L.V.~Laperashvili, Int.J.Mod.Phys. A {\bf 20}, 5911
(2005).
\bibitem{18}
D.L.~Bennett, L.V.~Laperashvili, H.B.~Nielsen, {\it Relation
between fine structure constants at the Planck scale from multiple
point principle,} in: {\it  Proceedings to the 9th Workshop: What
comes beyond the standard models,} Bled, Slovenia (DMFA,
Zaloznistvo, Ljubljana, M.~Breskvar et al., Dec 2006), p.10;
arXiv: hep--ph/0612250.
\bibitem{19}
D.L.~Bennett, L.V.~Laperashvili, H.B.~Nielsen, {\it Finestructure
constants at the Planck scale from multiple point principle,} in:
{\it  Proceedings to the 10th Workshop on What Comes Beyond the
Standard Model,} Bled, Slovenia, 17-27 Jul 2007; arXiv: 0711.4681
[hep-ph].
\bibitem{20}
J.~Conway et al. {\it Search for Heavy Top-like Quarks Using
Lepton Plus Jets Events in 1.96-TeV p anti-p Collisions.} CDF
Collaboration FERMILAB-PUB-08-017-E, Jan 2008; arXiv: 0801.3877
[hep-ex].
\bibitem{21}
J.~Convay et
al.\\http://www-cdf.fnal.gov/physics/new/top/2008/tprop/public.html.
\bibitem{22}
E.K.G.~Sarkisyan, A.S.~Sakharov. {\it Multihadron production
features in different reactions.} Invited talk at 35th
International Symposium on Multiparticle Dynamics (ISMD 05),
Kromeriz, Czech Republic, 9-15 Aug 2005. Published in: AIP
Conf.Proc. {\bf 828} 35-41, 2006. Also in: *Kromeriz 2005,
Multiparticle dynamics*, 35-41; arXiv: hep-ph/0510191.

\end{thebibliography}
\end{document}